\documentclass[journal,draftclsnofoot,onecolumn,12pt]{IEEEtranTCOM}
\usepackage{array,cite}
\usepackage{graphicx,epsfig,subfigure}
\usepackage{amsthm}
\usepackage{amsmath, nccmath}
\usepackage{mdwmath}
\usepackage{array}
\usepackage{subeqnarray}
\usepackage{stfloats}
\usepackage{algorithm}
\usepackage{xcolor}
\usepackage{amssymb}

\usepackage{algpseudocode}

\usepackage{tabularx}
\usepackage{booktabs}

\begin{document}

\title{\renewcommand{\baselinestretch}{1.2} \Huge
MmWave Communication With Active Ambient Perception}
\author{\normalsize Chunxu~Jiao, Zhaoyang~Zhang$^{\dagger}$, Caijun~Zhong, Xiaoming~Chen, and Zhiyong~Feng
\thanks{Part of this work was presented at IEEE ICC 2018 \cite{jiao2018mmWave}.}
\thanks{Chunxu~Jiao (Email: {\tt jiaocx1990@zju.edu.cn}), Zhaoyang~Zhang (Corresponding Author, Email: {\tt ning\_ming@zju.edu.cn}), Caijun~Zhong (Email: {\tt caijunzhong@zju.edu.cn}) and Xiaoming~Chen (Email: {\tt chen\_xiaoming@zju.edu.cn}) are with the College of Information Science and Electronic Engineering, Zhejiang University, Hangzhou, China.}
\thanks{Zhiyong~Feng (Email: {\tt fengzy@bupt.edu.cn}) is with Beijing University of Posts and Telecommunications, Beijing, China.}
}
\maketitle
\vspace{-2cm}
\begin{abstract}
In existing communication systems, the channel state information of each UE (user equipment) should be repeatedly estimated when it moves to a new position or another UE takes its place. The underlying ambient information, including the specific layout of potential reflectors, which provides more detailed information about all UEs' channel structures, has not been fully explored and exploited. In this paper, we rethink the mmWave channel estimation problem in a new and indirect way, i.e., instead of estimating the resultant composite channel response at each time and for any specific location, we first conduct the ambient perception exploiting the fascinating radar capability of a mmWave antenna array and then accomplish the location-based sparse channel reconstruction. In this way, the sparse channel for a quasi-static UE arriving at a specific location can be rapidly synthesized based on the perceived ambient information, thus greatly reducing the signalling overhead and online computational complexity. Based on the reconstructed mmWave channel, single-beam mmWave communication is designed and evaluated which shows excellent performance. Such an approach in fact integrates radar with communication, which may possibly open a new paradigm for future communication system design.
\end{abstract}
\vspace{-0.5cm}

\section{Introduction}



In recent years, millimeter-wave (mmWave) has attracted lots of attention due to narrow and highly-directional transmission beams. To reap its full benefits, transmit and receive beamformers must be adjusted to match the channel's angle of departures (AoDs) and angle of arrivals (AoAs). Considering the mobility of user equipment (UE), these angles are possible to vary continuously. To tackle this issue, an intuitive solution is to perform channel estimation repeatedly. However, this results in high estimation overhead and long channel state information (CSI) delay. As such, how to acquire the ever-changing channel efficiently and promptly has been a major consideration in mmWave communication.

The popular recipe for channel estimation mainly exploits the sparse nature of the mmWave channel, which enables a class of schemes based on compressed sensing and proper codebook design \cite{alkhateeb2014channel,singh2015on,xie2017unified}. Such schemes usually provide channel estimate for some specific time and/or location in forms of quantized AoD and AoA, etc. To get rid of the quantized angle assumption, \cite{zhu2017hybrid} estimated the AoA spectrum of the channel through beam scanning. Although these schemes dramatically improve the estimation efficiency in one snapshot, they have to repeat the estimation process when the UE moves to a new position or another UE takes its place, even if multiple UEs are sharing the same ambient environment. However, as well shown in the popular ray-tracing channel modeling \cite{tam1995propagation,athanasiadou2000novel}, the underlying ambient information, including the physical layout of potential scatters and reflectors, in fact provides rather detailed information about UEs' channel structures. In particular, as noted in \cite{kaya2009characterizing}, for indoor areas where the layouts and materials of walls, floors and furnishings can be known \emph{a priori}, the mmWave channel can be reconstructed on a very large scale without too much effort. As for outdoor scenarios, it was confirmed in \cite{vasisht2016eliminating} that channel estimation might be unnecessary if rays are traced beforehand. \cite{han2018efficient} further utilized limited feedback to refine the tracing results. In a recent study \cite{alkhateeb2019deepmimo}, a parameterized ray-tracing channel dataset is generated by machine learning tools and shows promising capabilities in modelling indoor and outdoor mmWave MIMO channels.

The above observations motivate us to rethink the mmWave channel estimation problem in a new and indirect way, i.e., instead of estimating the resultant composite channel response at each time and for any specific location, we first conduct the ambient sensing and perception exploiting the fascinating radar capability of mmWave antenna array and then accomplish the location-based sparse channel reconstruction with ray-tracing method. More specifically, frequency modulated continuous wave (FMCW) radar \cite{schuster2006performance,pauli2017miniaturized,adib20143d}, which derives ranging results from the frequency variation caused by propagation, is applied to sense the environment. According to a designed perception criteria, the layouts and materials of interior objects are determined. Then, based on the perceived ambient information, the CSI for a static or slowly moving user can be rapidly synthesized via ray-tracing, and even possibly reused by other users arriving at the same location, with a greatly reduced signalling overhead and online computational complexity. In addition, the unique channel characteristics of mmWave facilitate the channel reconstruction procedure. On the one hand, mmWave propagation channel is quasi-optical in nature \cite{maltsev2009experimental}, which conforms to the geometrical optics assumption in ray-tracing. On the other hand, mmWave communication link is predominately composed of one line-of-sight (LoS) ray and just a few non-line-of-sight (NLoS) reflected rays, thus confining the overall computational overhead. Notably, such an approach in fact integrates radar with communication, which may possibly open a new paradigm for communication system design.

In this paper, we consider an indoor mmWave beamforming communication system with a FMCW radar mode to accomplish the above ambient perception and channel reconstruction. Note that with a similar idea, J. Chio \emph{et al.} proposed a radar-aided beamforming approach in \cite{choi2016millimeter}. However, their sensing and transceiving are realized via two separate arrays instead of a combined system. In our work, the mmWave access point (AP) first senses and perceives the indoor ambient environemnt using the radar mode. Then based on the perceived ambient information, the propagation channel from the AP to a certain UE is reconstructed efficiently and promptly, based on which the communication thus takes place.

The main contributions of our work can be summarized as follows:

\begin{itemize}
    \item An indoor ambient sensing scheme in mmWave band is presented to support AP's radar mode. FMCW is exploited to acquire the range information. An innovative mmWave reflection model is proposed to analyze the echo signals, based on which the material information can also be derived.
    \item The ambient perception procedure is presented. With ambient perceiving and ray-tracing, a mmWave channel reconstruction scheme is proposed. The reconstructed channel contains the AoD, AoA and gain of each propagation path. The performance and robustness of the proposed scheme are discussed and examined through numerical simulations.
    \item Using the reconstructed mmWave channel, a single-beam communication link is established along the strongest path from the AP to a certain UE. Through comparing the rates produced by the reconstructed channel and the actual channel, the good performance of the proposed scheme is demonstrated.
\end{itemize}

The remainder is organized as below. Section II describes the ambient sensing scheme. In Section III, the ambient perception and the subsequent channel reconstruction procedures are presented. Next, Section IV derives the rate of single-beam communication based on the proposed scheme. Numerical results are given in Section V. And finally, Section VI concludes the work.

Throughout rest of this macuscript, matrices and vectors are denoted by bold capital and lowercase letters, respectively. $\textrm{U}\left(a,b\right)$ represents uniform distribution over $(a, b)$. $\left( \cdot  \right)^{T}$, $\left( \cdot  \right)^{H}$, $\textrm{mod}(\cdot,\cdot)$, $\mathbb{E}\left[\cdot\right]$ represent the transpose, Hermitian transpose, modulo and expectation operations, $\left|\cdot\right|$ calculates the absolute value, $\mathcal{N}\left(0,\sigma^{2}\right)$ is Gaussian distribution with mean $0$ and variance $\sigma^{2}$, and $\mathcal{CN}\left(\textbf{0},\sigma^{2}\textbf{I}\right)$ refers to complex Gaussian distribution with mean $\textbf{0}$ and covariance matrix $\sigma^{2}\textbf{I}$.



\section{Ambient Sensing}

\begin{figure}[!t]\centering
\includegraphics[angle=0,scale=0.19]{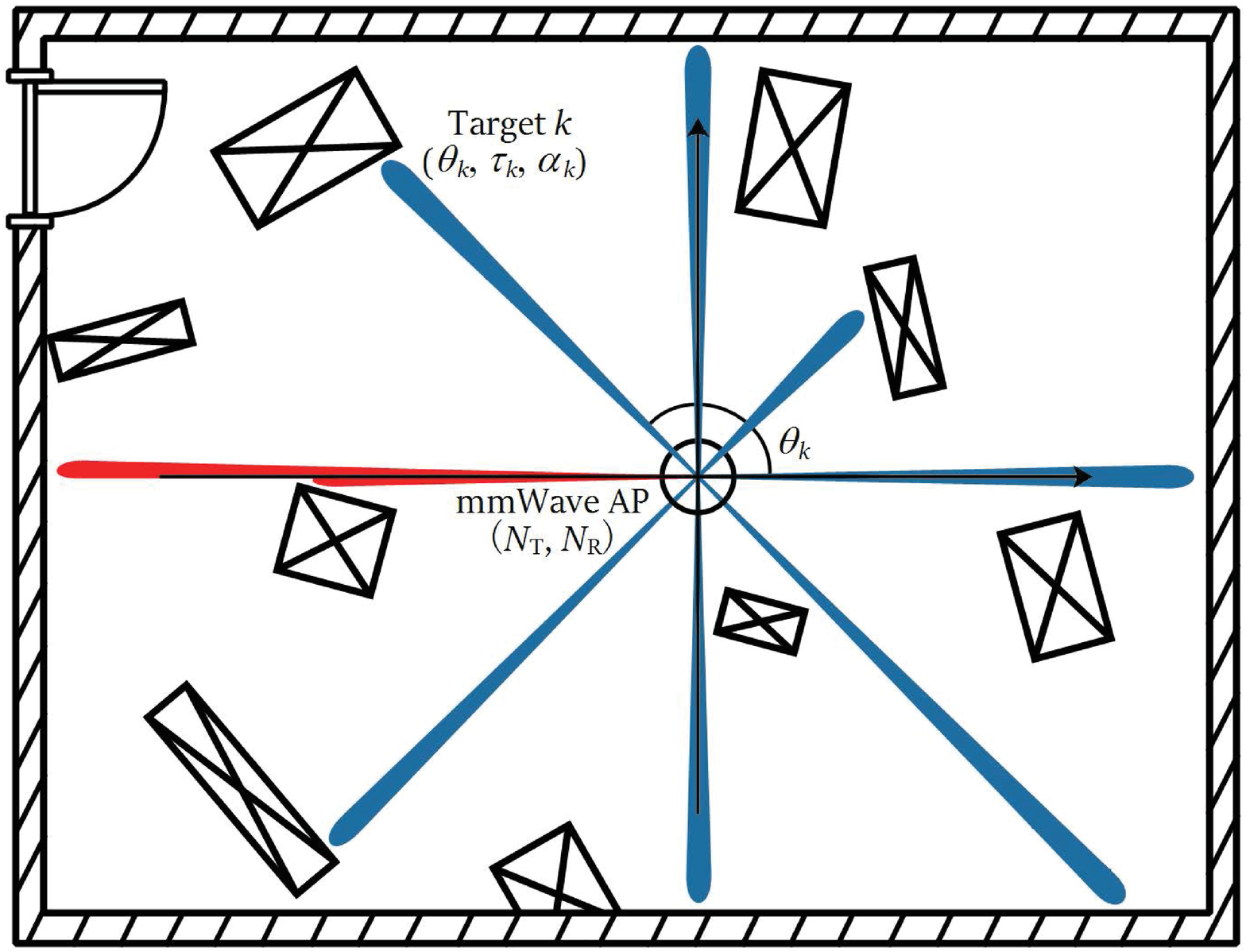}
\vspace{-0.675cm}
\caption{MmWave radar system model: the room is treated as $K$ targets (reflection points), in which the $k$-th target is characterized by its azimuth $\theta_{k}$, delay $\tau_{k}$ and reflection coefficient $\alpha_{k}$.}
\label{Fig_sensing_model}
\vspace{-0.7cm}
\end{figure}

We consider a two-dimensional (2-D) indoor scenario as illustrated in Fig. \ref{Fig_sensing_model}. A mmWave AP with collocated transmitter and receiver is mounted in the room. The transmit and receive arrays are uniform circular arrays (UCA) and comprise $N_{\textrm{T}}$ transmit and $N_{\textrm{R}}$ receive antennas, respectively. Both arrays have massive antennas to form narrow beams. Due to the short wavelength of mmWave signals, the array size remains small and practical. In regard of the ambient obstacles including furnitures, appliances and walls in the room, we approximate it with a series of random objects that form a Boolean scheme of rectangles \cite{bai2014analysis}.

\subsection{Indoor Scenario}\label{indoor_scenario}

Let $L_{\textrm{r}}$ and $W_{\textrm{r}}$ be the length and width of the room. Considering the AP's central position as the origin, a Cartesian coordinate system can be simply established. Invoking this setting, some further assumptions are made to clarify the indoor scenario.

\emph{Assumption 1:} The AP and objects on the 2-D transverse plane are at the same high level.

\emph{Assumption 2:} The centers of objects are generated using poisson point process with normalized mean parameter $\lambda_{\textrm{PPP}}$ \cite{bai2014analysis}. The average number of objects is $\lambda_{\textrm{PPP}}L_{\textrm{r}}W_{\textrm{r}}$.

\emph{Assumption 3:} Indoor objects are rectangular with flat surfaces \cite{piesiewicz2008performance,priebe2013stochastic}.

\emph{Assumption 4:} The length and width of each object conform to $\textrm{U}\left(0,L_{\textrm{o}}\right)$ and $\textrm{U}\left(0,W_{\textrm{o}}\right)$, respectively. Objects' tilt angles are generated based on $\textrm{U}\left(0,\pi\right)$.

\emph{Assumption 5:} The properties, especially those related to reflectance, of all possible materials in the room are known \emph{a priori}.

\emph{Remarks:} Assumption 3 is practical since indoor furnishings and walls are usually regularly shaped. Assumption 5 is practical because sufficient off-line measurements can be conducted beforehand.

\subsection{Sensing with FMCW Radar}\label{FMCW}



To sense such ambient environment, the mmWave AP sequentially \emph{samples} the interior space with $K$ probing beams. The $k$-th beam contains a vector $\textbf{x}_{k}\left(t\right)\in \mathbb{C}^{N_{\textrm{T}}\times 1}$ and is steered towards the direction $\theta_{k}\triangleq\frac{k}{K}2\pi$. The corresponding half power beamwidth $W\left(\theta_{k},N_{\textrm{T}}\right)$ is calculated as the range of $\theta$ that satisfies
\begin{equation}
\left|\textbf{a}_{\textrm{T}}^{H}\left(\theta_{k},N_{\textrm{T}}\right)\textbf{a}_{\textrm{T}}\left(\theta,N_{\textrm{T}}\right)/N_{\textrm{T}}\right|^2\geq\frac{1}{2},
\label{bwidth}
\end{equation}
where $\textbf{a}_{\textrm{T}}\left(\theta,N_{\textrm{T}}\right)\in \mathbb{C}^{N_{\textrm{T}}\times 1}$ denotes the steering vector at the transmit array and is obtained by
\begin{equation}
\textbf{a}_{\textrm{T}}\left(\theta,N_{\textrm{T}}\right)=\left[e^{j \frac{2\pi}{\lambda} r \cos(\theta)},e^{j \frac{2\pi}{\lambda} r \cos(\theta-\frac{1}{N_{\textrm{T}}}2\pi)},\cdots,e^{j \frac{2\pi}{\lambda} r \cos(\theta-\frac{N_{\textrm{T}}-1}{N_{\textrm{T}}}2\pi)}\right]^{T},
\label{steering_vector_n}
\end{equation}
$\lambda$ is the wavelength, and $r$ refers to the radius of the circular array. Notably, UCA is preferred in this scenario because it provides nearly identical beamwidth for every direction $\{\theta_{k}\}_{k=1,\cdots,K}$. Similar to \eqref{steering_vector_n}, the receive steering vector $\textbf{a}_{\textrm{R}}\left(\theta,N_{\textrm{R}}\right)\in \mathbb{C}^{N_{\textrm{R}}\times 1}$ is
\begin{equation}
\textbf{a}_{\textrm{R}}\left(\theta,N_{\textrm{R}}\right)=\left[e^{j \frac{2\pi}{\lambda} r \cos(\theta)},e^{j \frac{2\pi}{\lambda} r \cos(\theta-\frac{1}{N_{\textrm{R}}}2\pi)},\cdots,e^{j \frac{2\pi}{\lambda} r \cos(\theta-\frac{N_{\textrm{R}}-1}{N_{\textrm{R}}}2\pi)}\right]^{T}.
\label{steering_vector_n_Re}
\end{equation}

When the $k$-th beam encounters an object or the wall, it is partially reflected back to the AP. From a radar point of view, the corresponding \emph{reflection point} is regarded as a target, denoted by $\textrm{T}_{k}$ for ease of exposition, and can be characterized by a propagation delay $\tau_{k}$ and a complex reflection coefficient $\alpha_{k}$. On these basis, the received vector at the AP is expressed as \cite{heckel2016super}
\begin{equation}
\textbf{y}_{k}\left(t\right)=\alpha_{k}\textbf{a}_{\textrm{R}}\left(\theta_{k},N_{\textrm{R}}\right)\textbf{a}_{\textrm{T}}^{H}\left(\theta_{k},N_{\textrm{T}}\right)\textbf{x}_{k}\left(t-2\tau_{k}\right).
\label{receive_signal}
\end{equation}
Then, it is needed to derive $\tau_{k}$ and $\alpha_{k}$ from the response $\textbf{y}_{k}\left(t\right)\in \mathbb{C}^{N_{\textrm{R}}\times 1}$. To address this issue, we appeal to a commonly-used ranging technique called FMCW radar. Its basic idea is to multiply a frequency modulated signal $s\left(t\right)$ by its echo $z\left(t\right)=\alpha s\left(t-2\tau\right)$ and derive the \emph{beat frequency}, which has a mathematical relationship with $\tau$. For better clarity, some details are given below.

First, the transmit signal $\textbf{x}_{k}$ is
\begin{equation}
\textbf{x}_{k}\left(t\right)=\sqrt{P_{\textrm{T}}}\textbf{f}_{\textrm{RF},k} s\left(t\right),
\label{transmit_signal_vector_pro}
\end{equation}
where $P_{\textrm{T}}$ denotes the transmit power of each antenna, $\textbf{f}_{\textrm{RF},k}\in \mathbb{C}^{N_\textrm{T}\times 1}$ is the analog beamformer, and the frequency modulated $s\left(t\right)$ is expressed as \cite{schuster2006performance}
\begin{equation}
s\left(t\right)=e^{j2\pi\int_{0}^{t}(f_{\textrm{c}}+f'\tau)d\tau}=e^{j2\pi\left(f_{\textrm{c}}+\frac{f'}{2}t\right)t},
\label{transmit_signal_FMCW}
\end{equation}
$f_{\textrm{c}}$ is the initial carrier frequency, $f'$ is a given chirp rate of frequency sweeping. The ultra-high frequency band gives rise to the fact $f't\ll f_{\textrm{c}}$. Consequently, the wavelength fluctuation incurred by frequency sweeping is ignored.

Substituting (\ref{transmit_signal_vector_pro}) and (\ref{transmit_signal_FMCW}) into (\ref{receive_signal}), the received FMCW signal is obtained by
\begin{equation}
\textbf{y}_{k}\left(t\right)=\alpha_{k}\sqrt{P_{\textrm{T}}}\textbf{a}_{\textrm{R}}\left(\theta_{k},N_{\textrm{T}}\right)\textbf{a}_{\textrm{T}}^{H}\left(\theta_{k},N_{\textrm{T}}\right)\textbf{f}_{\textrm{RF},k} e^{j2\pi\left(f_{\textrm{c}}+\frac{f'}{2}\left(t-2\tau_{k}\right)\right)\left(t-2\tau_{k}\right)}.
\label{receive_signal_pro}
\end{equation}
Without loss of generality, we exploit simple transmit and receive beamformers as follows,
\begin{equation}
\textbf{f}_{\textrm{RF},k}=\frac{\textbf{a}_{\textrm{T}}\left(\theta_{k},N_{\textrm{T}}\right)}{\sqrt{N_{\textrm{T}}}}, ~\mbox{ and }~
\textbf{w}_{\textrm{RF},k}=\frac{\textbf{a}_{\textrm{R}}^{H}\left(\theta_{k},N_{\textrm{T}}\right)}{\sqrt{N_{\textrm{R}}}}. \label{beamformer}
\end{equation}
Then the post-processed FMCW signal is
\begin{equation}
z_{k}\left(t\right)=\textbf{w}_{\textrm{RF},k}\textbf{y}_{k}\left(t\right)=\alpha_{k} \sqrt{P_{\textrm{T}}N_{\textrm{T}}N_{\textrm{R}}} e^{j2\pi\left(f_{\textrm{c}}+\frac{f'}{2}\left(t-2\tau_{k}\right)\right)\left(t-2\tau_{k}\right)}.
\label{receive_signal_pro2}
\end{equation}
This signal is then multiplied by $s\left(t\right)$ and low-pass filtered, thus resulting in the mixed signal
\begin{equation}
m_{k}\left(t\right)=\alpha_{k} \sqrt{P_{\textrm{T}}N_{\textrm{T}}N_{\textrm{R}}} e^{-j2\pi \left( 2f' \tau_{k}t + 2 f_{\textrm{c}}\tau_{k} - 2 f' \tau_{k}^{2}\right)}.
\label{mix_signal}
\end{equation}
The carrier frequency of $m_{k}\left(t\right)$, i.e., $2f' \tau_{k}$, is commonly referred to as beat frequency and can be measured accurately based on properly configured sampling rate and fast fourier transform (FFT) size. Invoking the measured result $\hat{f}_{\textrm{b},k}$, the estimated delay $\hat{\tau}_{k}$ can be simply derived as
\begin{equation}
\hat{\tau}_{k} = \hat{f}_{\textrm{b},k}/\left(2 f'\right).
\label{delay}
\end{equation}

With a closer look at \eqref{receive_signal_pro2} and \eqref{mix_signal}, we see the auto-correlation of $z_{k}\left(t\right)$ or $m_{k}\left(t\right)$ equals $P_{\textrm{T}}N_{\textrm{T}}N_{\textrm{R}}\left| \alpha_{k} \right|^{2}$. As such, $\left| \hat{\alpha}_{k} \right|$, i.e., the estimated amplitude of $\alpha_{k}$, can be obtained. Yet the phase is still not clear. To tackle this issue, a more detailed model of the reflection coefficient $\alpha_{k}$ is needed. This model is elaborated further in the next subsection.

It is worth noting that there might be special cases that $\tau_{k}$ and $\alpha_{k}$ are erroneously measured, see the red beam that meets multiple objects in Fig. \ref{Fig_sensing_model} for illustration. Under this condition, the received signal becomes a superposition of several echoes with different delays and reflection coefficients, making the estimations rather complicated. For ease of analysis, we discard these results and set $\hat{\tau}_{k} = \infty$, $\left|\hat{\alpha}_{k}\right|=0$. In addition, as the focus of this paper is not on measuring accuracy, we assume all the remaining measurements are precise. As such, the hats in $\{\hat{\tau}_{k}\}_{k=1,\cdots,K}$ and $\{\left| \hat{\alpha}_{k} \right|\}_{k=1,\cdots,K}$ are dropped for notational convenience.

\subsection{Reflection Model}\label{reflection_model}

\begin{figure}[H]
\centering
\subfigure[At reflection point $O$, the material related reflectance is $(R_{\textrm{s}},R_{\textrm{d}})$. The directions pointing towards the transmit and receive array are $\varphi_{\textrm{in}}$ and $\varphi_{\textrm{R}}$. The corresponding distances are $D_{\textrm{T}}$ and $D_{\textrm{R}}$.]{\label{Fig_reflectance_model_a}
\includegraphics[angle=0,scale=0.18]{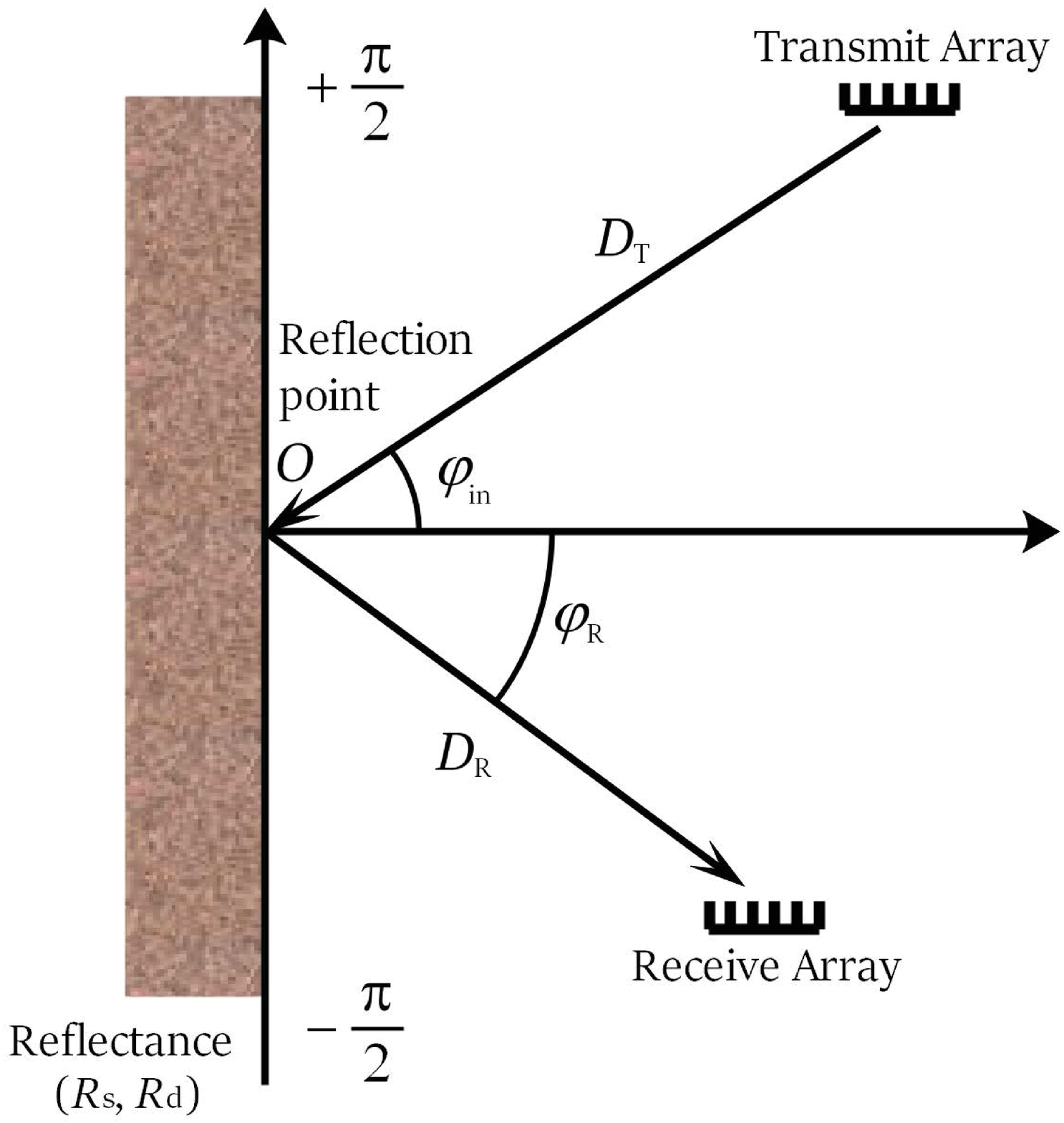}}
\hspace{1cm}
\subfigure[The reflected wave involves specular (blue solid line) and diffuse reflection (red dotted line). Specular reflection points to direction $\varphi_{\textrm{re}}$. Diffuse reflection radiates towards the whole space with a specific power density (red circle).]{\label{Fig_reflectance_model_b}
\includegraphics[angle=0,scale=0.18]{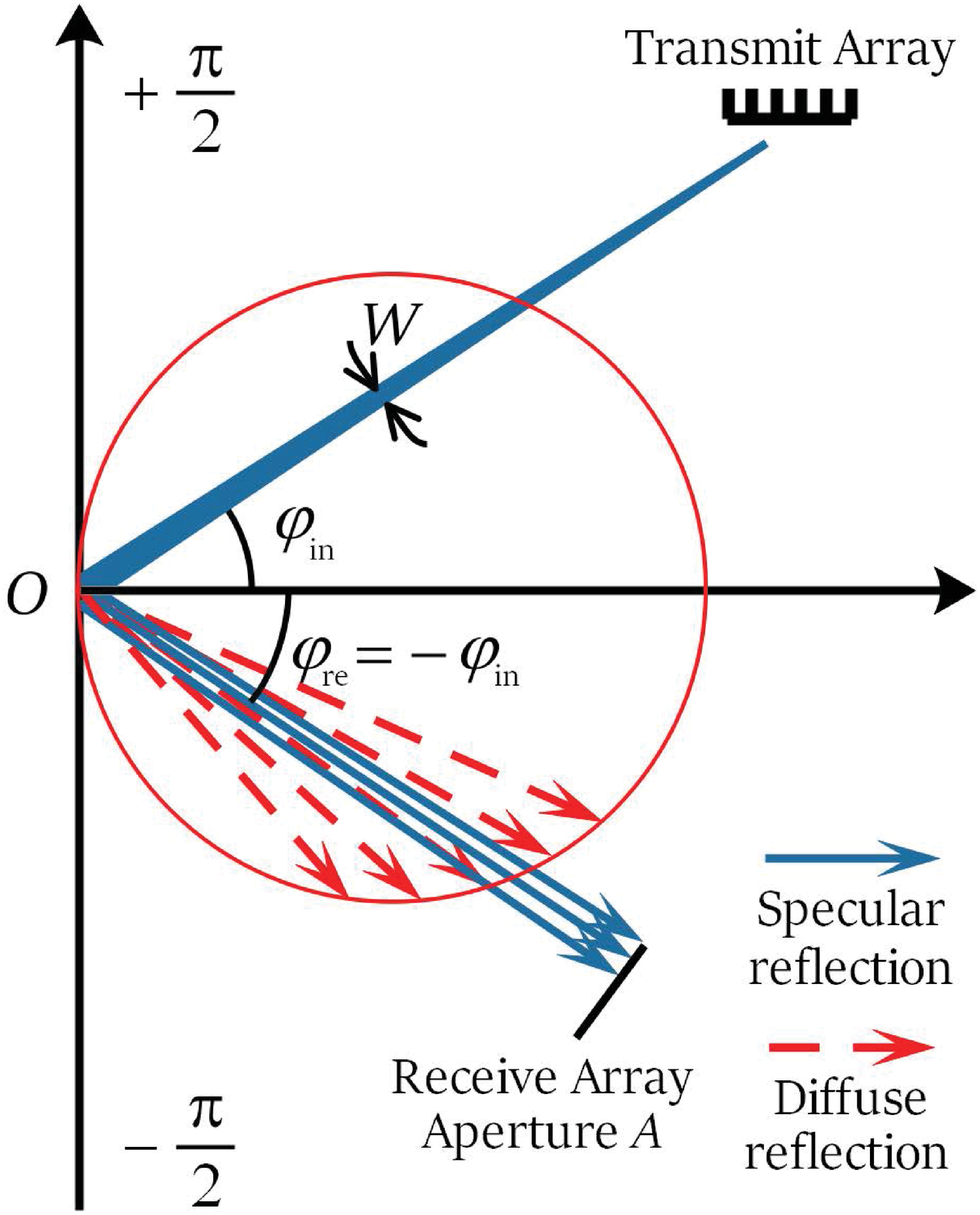}}
\caption{Generalized MmWave reflection model. The transmit and receive array may be non-collocated.}
\label{Fig_reflectance_model}
\vspace{-1.0cm}
\end{figure}

Different from sub-6GHz band waves, mmWave propagation exhibits quasi-optical behavior. As such, we propose to describe the reflected rays with the Phong reflection model \cite{phong1975illumination}. For a better understanding, consider a generalized case in Fig. \ref{Fig_reflectance_model}, where a mmWave transmitter sends signals to a receiver through single-bounce reflection. Related parameters are summarized in TABLE I. As shown in Fig. \ref{Fig_reflectance_model_a}, the transmit wave hits the material surface at point $O$ and is partially reflected to the receive array. The directions pointing towards the transmit and receive array are $\varphi_{\textrm{in}}$ and $\varphi_{\textrm{R}}$, respectively. The corresponding distances are $D_{\textrm{T}}$ and $D_{\textrm{R}}$, respectively. The radar cross-section (RCS) parameters of the material surface includes the reflectance $(R_{\textrm{s}},R_{\textrm{d}})$ and the ratio $\eta$ that describes the absorptivity. Based on the Phong reflection model, the reflected wave involves specular and diffuse reflection, see Fig. \ref{Fig_reflectance_model_b} for illustration. As the specular counterpart is highly directional, the induced illumination is restricted to a small region in direction $\varphi_{\textrm{re}}=-\varphi_{\textrm{in}}$. The area of illumination depends on the beamwidth $W$ and the propagation distance $D_{\textrm{T}}+D_{\textrm{R}}$. By contrast, the diffuse counterpart radiates towards the whole angular range, i.e., $(-\frac{\pi}{2},\frac{\pi}{2})$. These two reflection components co-determine the reflection coefficient at point $O$ through the following proposition.

\begin{table}[H]
\renewcommand\arraystretch{0.6}
\begin{center}
\caption[]{Reflection Model Parameters}
\begin{tabular}{|c|c|}
\hline
$R_{\textrm{s}}$&Specular reflectance, $R_{\textrm{s}}\in (0,1)$\\ \hline
$R_{\textrm{d}}$&Diffuse reflectance, $R_{\textrm{d}}\in (0,1)$\\ \hline
$\varphi_{\textrm{in}}$&Incident angle of the mmWave beam  \\ \hline
$\varphi_{\textrm{re}}$&Reflected angle of the mmWave beam, $\varphi_{\textrm{re}}=-\varphi_{\textrm{in}}$ \\ \hline
$\varphi_{\textrm{R}}$&Direction pointing towards the receive array\\ \hline
$W$&Width of the transmit beam\\ \hline
$\omega$&Angular range that is covered by the receive array\\ \hline
$A$&2-D aperture of the receive array\\ \hline
$D_{\textrm{T}}$&Distance from the reflection point to the transmit array\\ \hline
$D_{\textrm{R}}$&Distance from the reflection point to the receive array \\ \hline
$\eta$&Ratio between the reflected power and the incident power, $\eta \in \left(0,1\right)$ \\ \hline
\end{tabular}
\end{center}
\vspace{-1.5cm}
\end{table}

\emph{Proposition 1:} The complex reflection coefficient $\alpha$ can be modeled by
\begin{align}
&\medmath{\alpha =
e^{-j(\mathit{\Phi}_{\textrm{re}}-\mathit{\Phi}_{\textrm{prop}})}\times} \notag\\
&\medmath{\sqrt{\eta \left(\frac{D_{2}-\max\left(\tan\left(\min\left(\left|\varphi_{\textrm{in}}+\varphi_{\textrm{R}}\right|,\arctan(D_{2})\right)\right),\left|D_{1}\right|\right)}{D_{2}-D_{1}} R_{\textrm{s}}+\cos\left(\varphi_{\textrm{in}}\right)\cos\left(\varphi_{\textrm{R}}\right)\frac{A}{\sqrt{4D_{\textrm{R}}^{2}+A^{2}}} R_{\textrm{d}}\right)}},
\label{reflection_coefficient}
\end{align}
where $D_{1} \triangleq \frac{A-W\left( D_{\textrm{T}} + D_{\textrm{R}} \right)}{2D_{\textrm{R}}}$, $D_{2} \triangleq \frac{A+W\left( D_{\textrm{T}} + D_{\textrm{R}} \right)}{2D_{\textrm{R}}}$, and $\mathit{\Phi}_{\textrm{re}}\in\left(0,2\pi\right)$ denotes the phase shift determined by the reflection property, and $\mathit{\Phi}_{\textrm{prop}}=\frac{2\pi(D_{\textrm{T}}+D_{\textrm{R}})}{\lambda}$ is the phase shift induced by propagation.

\begin{IEEEproof}
This proposition is obtained through the following two steps. First, let us look at the specular counterpart. As depicted in Fig. \ref{Fig_reflectance_model}, the specular reflection is mainly impacted by three parameters: a) $\varphi_{\textrm{re}}$ determines the orientation; b) $R_{\textrm{s}}$ characterizes the strength; c) $W$ impacts the illumination region. In the meantime, the specular reflection ray can be received if and only if the receive array aperture $A$ overlaps with the illumination region. Recall that mmWave array is usually large, $W$ tends to be quite small. Therefore, the illumination region can be approximated by a line segment $W(D_{\textrm{T}}+D_{\textrm{R}})$ that is perpendicular to direction $\varphi_{\textrm{re}}$. Through some geometric analyses, the condition that the receive aperture overlaps with the illumination area is
\begin{equation}
\tan(\left|\varphi_{\textrm{re}}-\varphi_{\textrm{R}}\right|)<\frac{A+W(D_{\textrm{T}}+D_{\textrm{R}})}{2D_{\textrm{R}}}.
\end{equation}
To facilitate understanding, we show the special case that $\tan(\varphi_{\textrm{re}}-\varphi_{\textrm{R}})=\frac{A+W(D_{\textrm{T}}+D_{\textrm{R}})}{2D_{\textrm{R}}}$ in Fig. \ref{Fig_proposition1}. Due to symmetry, the conclusion above is obtained.

\begin{figure}[ht]\centering
\includegraphics[angle=0,scale=0.15]{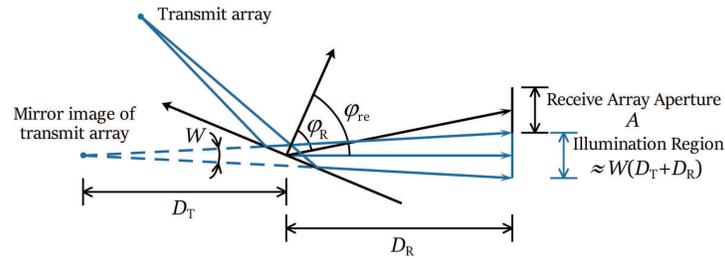}
\vspace{-0.5cm}
\caption{Special case that the receive array aperture does not overlap with the illumination region. In this case, no specular reflection can be received.}
\label{Fig_proposition1}
\vspace{-0.5cm}
\end{figure}

Let $P_{\textrm{in}}$ be the incident power at the reflection point $O$. Define $D_{1} \triangleq \frac{A-W\left( D_{\textrm{T}} + D_{\textrm{R}} \right)}{2D_{\textrm{R}}}$ and $D_{2} \triangleq \frac{A+W\left( D_{\textrm{T}} + D_{\textrm{R}} \right)}{2D_{\textrm{R}}}$. Then, the overlapping situations can be divided into the following two cases:

Case 1: Full overlap, i.e., $\left|\varphi_{\textrm{re}}-\varphi_{\textrm{R}}\right|<\arctan\left|D_{1}\right|$.

In this case, the received power is $\eta R_{\textrm{s}} P_{\textrm{in}}$ if the array aperture is larger than the illumination region, and $\frac{A}{W\left( D_{\textrm{T}} + D_{\textrm{R}} \right)}\eta R_{\textrm{s}} P_{\textrm{in}}$ otherwise, where $\eta$ is the ratio between the reflected power and the incident power. In general, the specular reflected power at the receiver array is calculated as
\begin{equation}
P_{\textrm{re,s}}=\eta \left(\frac{D_{2}-\left|D_{1}\right|}{D_{2}-D_{1}} \right)R_{\textrm{s}} P_{\textrm{in}}.
\label{case_1}
\end{equation}

Case 2: Partial overlap, i.e., $\arctan\left|D_{1}\right|\leq\left|\varphi_{\textrm{re}}-\varphi_{\textrm{R}}\right|<\arctan (D_{2})$.

In this case, only a part of the receive array aperture overlaps with the illumination region. The overlapping part has the length $D_{\textrm{R}}(D_{2}-\tan\left|\varphi_{\textrm{re}}-\varphi_{\textrm{R}}\right|)$. Hence, the captured power can be obtained as
\begin{equation}
P_{\textrm{re,s}}=\eta\left(\frac{D_{\textrm{R}}(D_{2}-\tan\left|\varphi_{\textrm{re}}-\varphi_{\textrm{R}}\right|)}{W(D_{\textrm{T}} + D_{\textrm{R}})}\right)R_{\textrm{s}} P_{\textrm{in}}=\eta\left(\frac{D_{2}-\tan\left|\varphi_{\textrm{re}}-\varphi_{\textrm{R}}\right|}{D_{2}-D_{1}}\right)R_{\textrm{s}} P_{\textrm{in}}.
\label{case_2}
\end{equation}

Now combining Case 1 and Case 2, the captured specular reflection power is
\begin{equation}
P_{\textrm{re,s}}=\eta \left(\frac{D_{2}-\max\left(\tan\left(\min\left(\left|\varphi_{\textrm{in}}+\varphi_{\textrm{R}}\right|,\arctan(D_{2})\right)\right),\left|D_{1}\right|\right)}{D_{2}-D_{1}} \right)R_{\textrm{s}}P_{\textrm{in}}.
\label{specular_ref_power}
\end{equation}

Next, let us focus on the diffuse reflection. According to \cite{phong1975illumination} and Lambert¡¯s cosine law in optics, the observed diffuse reflection counterpart is directly proportional to $\cos\left(\varphi_{\textrm{in}}\right)$. This is equivalent to the condition that only the normal component of the incident beam is reflected, thus inducing the effective incident power $P_{\textrm{in}}\cos\left(\varphi_{\textrm{in}}\right)$. Such power is attenuated by power ratio $\eta$ and diffuse reflectance $R_{\textrm{d}}$, and then reflected to angular domain $\left(-\frac{\pi}{2},\frac{\pi}{2}\right)$. With the Phong reflection model, the angular power density function is given by
\begin{equation}
p_{\textrm{re,d}}\left(\varphi\right) = c_{\textrm{n}}\eta \cos\left(\varphi_{\textrm{in}}\right)\cos\left(\varphi\right) R_{\textrm{d}} P_{\textrm{in}},
\label{diffuse_ref_power_angular_density}
\end{equation}
where $c_{\textrm{n}}$ denotes the normalizing constant satisfying the law $\int_{-\frac{\pi}{2}}^{\frac{\pi}{2}}p_{\textrm{re,d}}\left(\varphi\right)d\varphi = \eta R_{\textrm{d}} P_{\textrm{in}}\cos\left(\varphi_{\textrm{in}}\right)$ and can be derived as $\frac{1}{2}$. Then, the diffuse reflection power can be obtained by
\begin{equation}
\label{diffuse_ref_power}
P_{\textrm{re,d}}
 = \int_{\varphi_{\textrm{R}}-\frac{1}{2}\omega}^{\varphi_{\textrm{R}}+\frac{1}{2}\omega}\frac{1}{2}\eta \cos\left(\varphi_{\textrm{in}}\right)\cos\left(\varphi\right) R_{\textrm{d}} P_{\textrm{in}} d\varphi = \eta \cos\left(\varphi_{\textrm{in}}\right)\cos\left(\varphi_{\textrm{R}}\right)\sin\left(\frac{\omega}{2}\right) R_{\textrm{d}} P_{\textrm{in}},
\end{equation}
where $\omega$ conforms to the relationship $\tan\left(\frac{\omega}{2}\right)=\frac{A}{2 D_{\textrm{R}}}$.

Invoking (\ref{specular_ref_power}) and (\ref{diffuse_ref_power}), the amplitude of the reflection coefficient is
\begin{align}
&\medmath{|\alpha|= \sqrt{\frac{P_{\textrm{re,s}}+P_{\textrm{re,d}}}{P_{\textrm{in}}}}} \notag \\
&\medmath{=\sqrt{\eta \left(\frac{D_{2}-\max\left(\tan\left(\min\left(\left|\varphi_{\textrm{in}}+\varphi_{\textrm{R}}\right|,\arctan(D_{2})\right)\right),\left|D_{1}\right|\right)}{D_{2}-D_{1}} R_{\textrm{s}}+\cos\left(\varphi_{\textrm{in}}\right)\cos\left(\varphi_{\textrm{R}}\right)\frac{A}{\sqrt{4D_{\textrm{R}}^{2}+A^{2}}} R_{\textrm{d}}\right)}}.
\end{align}
After applying the phase shifts $\mathit{\Phi}_{\textrm{re}}$ and $\mathit{\Phi}_{\textrm{prop}}$, the proof is complete.
\end{IEEEproof}

\emph{Remarks:} There are some intuitive observations from this proposition. The specular reflection counterpart can be considerably strong if the reflected beam points towards the receiver. This results from the highly-directional nature of specular reflection and lays a good foundation for NLoS transmission. Unfortunately, once the reflected beam is steered towards other directions, little power is captured. By contrast, the diffuse reflection counterpart is not reliant upon angular alignment and can be received invariably, which plays an important role in radar sensing. However, the diffuse reflection is severely impaired by the distance and might be too weak to support mmWave communication.

Based on this reflection model, the next section elaborates on the procedure for perceiving the ambient environment and reconstructing the mmWave channel. Specifically, ambient perception mainly focuses on diffuse reflection since the specular counterpart cannot be received in most cases, while channel reconstruction searches for specular rays to illuminate the receive array. Nonetheless, specular reflection is not precluded in ambient perception as there still exist special cases that the specular reflected beams are captured. Similarly, diffuse reflection is not precluded in channel reconstruction as it still has impacts upon the reflection coefficient.

\section{Ambient Perception and MmWave Channel Reconstruction}
In this section, we perform ambient perception using the delay information and the reflection coefficients derived in the sensing procedure. Then, based on the perceived ambient information, the mmWave channel is reconstructed with ray-tracing methodology.

\begin{figure}[!t]\centering
\includegraphics[angle=0,scale=0.24]{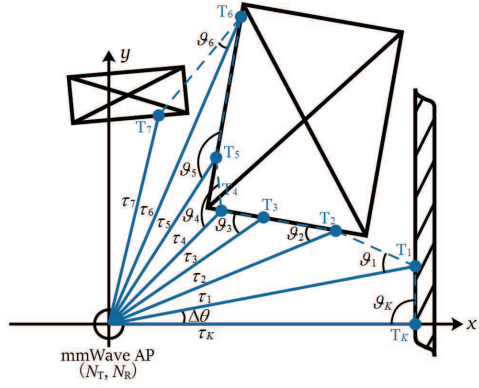}
\vspace{-0.5cm}
\caption{Illustration of target clustering.}
\label{Fig_target_clustering}
\vspace{-0.5cm}
\end{figure}

\begin{table}[!t]
\renewcommand\arraystretch{0.6}
\begin{center}
\caption[]{Target Clustering Parameters}
\begin{tabular}{|c|c|}
\hline
$\Delta \theta$&Beam angular spacing, $\Delta \theta = \frac{2\pi}{K}$\\ \hline
$A$&Array aperture of the UCA, $A=2 r$\\ \hline
$\vartheta_{k}$&Included angle between beam $k$ and the line defined by $\textrm{T}_{k}$ and $\textrm{T}_{k+1}$\footnotemark[1]  \\ \hline
\end{tabular}
\end{center}
\vspace{-1.5cm}
\end{table}

\footnotetext[1]{$\textrm{T}_{k+1}$ actually refers to $\textrm{T}_{\textrm{mod}(k,K)+1}$, meaning that $\vartheta_{K}$ corresponds to targets $\textrm{T}_{K}$ and $\textrm{T}_{1}$.}

\subsection{Ambient Perception}\label{target_clustering}

To perceive the mmWave ambient environment, two steps are required:
\begin{itemize}
    \item Target clustering. In this step, the mmWave AP divides the $K$ targets (flection points) into a number of clusters, each corresponding to a distinct flat surface.
    \item Surface forming. Given the target clusters, the AP forms a series of reflective surfaces with calculated layouts and reflection coefficients.
\end{itemize}
Further details are presented as below.

\subsubsection{Target clustering}

See the toy example in Fig. \ref{Fig_target_clustering}, where the first seven targets $\left\{ \textrm{T}_{1},\cdots,\textrm{T}_{7}\right\}$ and the last target $\textrm{T}_{K}$ are considered. Related parameters are given in TABLE II. In this situation, three clusters $\{\textrm{T}_{K},\textrm{T}_{1}\}$, $\{\textrm{T}_{2},\textrm{T}_{3},\textrm{T}_{4}\}$ and $\{\textrm{T}_{5},\textrm{T}_{6}\}$ are to be obtained from the measurements $\{\tau_{k}\}_{k=1,\cdots,K}$ and $\{\left| \alpha_{k} \right|\}_{k=1,\cdots,K}$. Another essential dataset is the reflection properties $\{( R_{\textrm{s}},R_{\textrm{d}},\mathit{\Phi}_{\textrm{re}})\}$ of all possible materials. Invoking the measurements in \cite{maltsev2009experimental,Jansen2008The,zhao201328}, such dataset can be known \emph{a priori}.

Now we are ready for target clustering. First, with the law of cosines, $\vartheta_{k}$ is calculated as
\begin{equation}
\vartheta_{k}=\arccos \left(\frac{\tau_{k}-\tau_{k+1}\cos\left(\Delta \theta\right)}{\sqrt{\tau_{k}^{2}+\tau_{k+1}^{2}-2\tau_{k}\tau_{k+1}\cos\left(\Delta \theta\right)}}\right).
\label{included_angle}
\end{equation}
Interestingly, when $\textrm{T}_{k}$ and $\textrm{T}_{k+1}$ are located on the same surface, $\vartheta_{k}$ and $\varphi_{\textrm{in},k}$ satisfy the equation $\varphi_{\textrm{in},k} = \vartheta_{k} - \frac{\pi}{2}$. In the meantime, since the transmit and receive array are collocated, it can be inferred from the definitions in TABLE I that $\varphi_{\textrm{R},k}=\varphi_{\textrm{in},k}$ and $D_{\textrm{R}}=D_{\textrm{T}}$. Substituting $A=2r$, $\varphi_{\textrm{R},k}=\varphi_{\textrm{in},k}$ and $D_{\textrm{R}}=D_{\textrm{T}}=c\tau_{k}$ into Proposition $1$, if $\textrm{T}_{k}$ and $\textrm{T}_{k+1}$ are on the same surface, then the reflection coefficient of $\textrm{T}_{k}$ can be modeled by
\begin{align}
&\medmath{\alpha_{k} =
e^{-j(\mathit{\Phi}_{\textrm{re}}-\mathit{\Phi}_{\textrm{prop}})}\times} \notag\\
&\medmath{~~~~~~\sqrt{\eta \left(\frac{D_{2,k}-\max\left(\tan\left(\min\left(\left|2\varphi_{\textrm{in},k}\right|,\arctan(D_{2,k})\right)\right),\left|D_{1,k}\right|\right)}{D_{2,k}-D_{1,k}} R_{\textrm{s}}+\cos^{2}\left(\varphi_{\textrm{in},k}\right)\frac{r}{\sqrt{(c\tau_{k})^{2}+r^{2}}} R_{\textrm{d}}\right)}} \notag \\
&\medmath{~~~\overset{(a)}{=}e^{-j(\mathit{\Phi}_{\textrm{re}}-\mathit{\Phi}_{\textrm{prop}})}\sqrt{\eta \left(\beta_{k} R_{\textrm{s}}+\gamma_{k} R_{\textrm{d}}\right)}},
\label{reflection_coefficient_pro}
\end{align}
where $\mathit{\Phi}_{\textrm{prop}}=\frac{4\pi c\tau_{k}}{\lambda}$, $D_{1,k} \triangleq \frac{r}{\textrm{c}\tau_{k}}-W(\theta_{k},N_{\textrm{T}})$, $D_{2,k} \triangleq \frac{r}{\textrm{c}\tau_{k}}+W(\theta_{k},N_{\textrm{T}})$, and $\textrm{c}$ represents the speed of light. Equation $(a)$ is derived through the definitions
\begin{equation}
\medmath{\beta_{k}\triangleq\frac{D_{2,k}-\max\left(\tan\left(\min\left(\left|2\varphi_{\textrm{in},k}\right|,\arctan(D_{2,k})\right)\right),\left|D_{1,k}\right|\right)}{D_{2,k}-D_{1,k}},~~~\gamma_{k}\triangleq\cos^{2}\left(\varphi_{\textrm{in},k}\right)\frac{r}{\sqrt{\left(\textrm{c}\tau_{k}\right)^{2}+r^{2}}}}.
\label{beta_gamma_k}
\end{equation}
Based on \eqref{reflection_coefficient_pro}, the reflection property $( \hat{R}_{\textrm{s},k},\hat{R}_{\textrm{d},k}, \hat{\mathit{\Phi}}_{\textrm{re},k})$ of $\textrm{T}_{k}$ can be estimated through
\begin{equation}
( \hat{R}_{\textrm{s},k},\hat{R}_{\textrm{d},k},\hat{\mathit{\Phi}}_{\textrm{re},k} )=\underset{(R_{1},R_{2},\mathit{\Phi})\in\{( R_{\textrm{s}},R_{\textrm{d}},\mathit{\Phi}_{\textrm{re}})\}}{\arg \min}| |\alpha_{k}|^{2} - \eta(\beta_{k} R_{1} + \gamma_{k} R_{2}) |.
\label{reflectance}
\end{equation}
By comparing the reflectance, when $( \hat{R}_{\textrm{s},k},\hat{R}_{\textrm{d},k} )=( \hat{R}_{\textrm{s},k+1},\hat{R}_{\textrm{d},k+1} )$ is observed, it can be inferred that $\textrm{T}_{k}$ and $\textrm{T}_{k+1}$ are located on the same surface and should be clustered. Nonetheless, there exist some situations that $R_{\textrm{s},k}$ and $R_{\textrm{d},k}$ are erroneously estimated yet $\textrm{T}_{k}$ is on the surface that contains $\textrm{T}_{k-1}$, such as $\textrm{T}_{4}$ in Fig. \ref{Fig_target_clustering}. To address this issue, the incident angle at $\textrm{T}_{4}$ is recalculated as $\varphi'_{\textrm{in},4}=\vartheta_{3}+\Delta \theta - \pi/2$. Then using (\ref{reflection_coefficient_pro}) and (\ref{reflectance}), a new reflectance estimation $( \hat{R}'_{\textrm{s},4},\hat{R}'_{\textrm{d},4} )$ is obtained. Provided that $( \hat{R}'_{\textrm{s},4},\hat{R}'_{\textrm{d},4} )=( \hat{R}_{\textrm{s},3},\hat{R}_{\textrm{d},3} )$, $\textrm{T}_{4}$ is included in cluster $\{ \textrm{T}_{2},\textrm{T}_{3} \}$. It is worth noting that the value of $( \hat{R}_{\textrm{s},4},\hat{R}_{\textrm{d},4} )$ should not be updated. This is because some targets may share identical reflectance but are located on different surfaces, e.g., $\textrm{T}_{4}$ and $\textrm{T}_{5}$.

To summarize, $\textrm{T}_{k}$ and $\textrm{T}_{k+1}$ are clustered together if the estimated reflectance $( \hat{R}_{\textrm{s},k},\hat{R}_{\textrm{d},k} )$ is equal to $( \hat{R}_{\textrm{s},k+1},\hat{R}_{\textrm{d},k+1} )$ or the updated version $( \hat{R}'_{\textrm{s},k+1},\hat{R}'_{\textrm{d},k+1} )$. The corresponding algorithm is presented in TABLE \ref{clustering_algorithm}. Related functions are presented in Appendix \ref{Functions}. Also, a predetermined threshold $\zeta$ is exploited to avoid miscalculation.

\begin{table}[!t]
\caption{Target Clustering Algorithm}\label{clustering_algorithm}\centering
\vspace{-1.2cm}
\begin{algorithm}[H]
\caption{Target Clustering}
\begin{algorithmic}[1]
\State Initialize the number of beams $K$, the lookup table $\{(R_{\textrm{s}},R_{\textrm{d}},\mathit{\Phi}_{\textrm{re}})\}$, the delay results $\{\tau_{k}\}$, the amplitudes $\{\left|\alpha_{k}\right|\}$, the inherent power ratio $\eta$, the aperture $A=2 r$, the beamwidth $\{W(\theta_{k},N_{\textrm{T}})\}$, the threshold value $\zeta$, set a cluster counter $l=0$, the included angles $\{\vartheta_{k}\}=\{0\}$, the reflectance at targets $\{(\hat{R}_{\textrm{s},k},\hat{R}_{\textrm{d},k})\}=\{(0,0)\}$, the clustering flags $\{f_{k}\}=\{0\}$;
\For {$k\in \left\{1,2,\cdots,K\right\}$}
\State Calculate $\vartheta_{k}$ with (\ref{included_angle});
\State $( \hat{R}_{\textrm{s},k},\hat{R}_{\textrm{d},k})=\textrm{REFLECTANCE}(k,\tau_{k},\vartheta_{k} - \pi/2)$;
\EndFor
\For {$k\in \left\{1,2,\cdots,K-1\right\}$}
    \State $f_{k+1}= \textrm{FLAG}(( \hat{R}_{\textrm{s},k},\hat{R}_{\textrm{d},k} ),( \hat{R}_{\textrm{s},k+1},\hat{R}_{\textrm{d},k+1} ))$;
    \State Expand or create clusters using CLUSTER$(l,f_{k},f_{k+1},( \hat{R}_{\textrm{s},k},\hat{R}_{\textrm{d},k} ))$;
\EndFor
\State Handle $T_{\textrm{K}}$, $T_{1}$ and the corresponding clusters with TAIL2HEAD$(l,f_{K},( \hat{R}_{\textrm{s},K},\hat{R}_{\textrm{d},K}),( \hat{R}_{\textrm{s},1},\hat{R}_{\textrm{d},1}))$;
\State Output each cluster and the corresponding reflectance;
\end{algorithmic}
\end{algorithm}
\vspace{-1.8cm}
\end{table}

We emphasize that the proposed algorithm provides good clustering results with hight probability. The reason is as follows. According to \eqref{reflectance}, a surface with reflectance $(R_{\textrm{s}},R_{\textrm{d}})$ is mistaken for $(R_{\textrm{s}}',R_{\textrm{d}}')$ only if
\begin{equation}
\left| \left(\beta R_{\textrm{s}} - \beta' R_{\textrm{s}}' \right) + \left(\gamma R_{\textrm{d}} - \gamma' R_{\textrm{d}}'\right) \right|<\zeta,
\label{error_clustering}
\end{equation}
where $\beta'$ and $\gamma'$ are calculated based on (\ref{beta_gamma_k}), while $\beta$ and $\gamma$ represent the actual values. As the transmitter and receiver are collocated, the specular reflection can rarely be captured. Therefore, let us first concentrate on the cases where $\beta=\beta'=0$. In this regard, (\ref{error_clustering}) can be recast as
\begin{equation}
\left| \gamma R_{\textrm{d}} -  \gamma' R_{\textrm{d}}' \right|<\zeta.
\label{error_clustering_pro}
\end{equation}
Invoking (\ref{beta_gamma_k}), this equation is equivalent to
\begin{equation}
\left| \cos^{2}\left(\varphi_{\textrm{in}}\right) R_{\textrm{d}} -  \cos^{2}\left(\varphi_{\textrm{in}}'\right) R_{\textrm{d}}' \right|<\zeta C,
\end{equation}
where $C\triangleq \frac{\sqrt{\left(\textrm{c}\tau_{k}\right)^{2}+r^{2}}}{r}$, $\varphi_{\textrm{in}}$ and $\varphi_{\textrm{in}}'$ are the actual and estimated incident angles, respectively. Then, the probability of \eqref{error_clustering_pro} can be obtained by
\begin{equation}
\medmath{P\left(\left| \gamma R_{\textrm{d}} -  \gamma' R_{\textrm{d}}' \right|<\zeta\right)=P\left(\sqrt{\frac{\cos^{2}\left(\varphi_{\textrm{in}}'\right) R_{\textrm{d}}'-\zeta C}{R_{\textrm{d}}}} < \cos\left(\varphi_{\textrm{in}}\right)<\sqrt{\frac{\cos^{2}\left(\varphi_{\textrm{in}}'\right) R_{\textrm{d}}'+\zeta C}{R_{\textrm{d}}}}\right).}
\label{PEP}
\end{equation}
It is observed that $P\left(\left| \gamma R_{\textrm{d}} -  \gamma' R_{\textrm{d}}' \right|<\zeta\right)\rightarrow 0$ as $\zeta \rightarrow 0$. Recall the pairwise error probability
\begin{align}
&\medmath{\textrm{PEP}\left((R_{\textrm{s}},R_{\textrm{d}})\rightarrow (R_{\textrm{s}}',R_{\textrm{d}}')\right)}\notag \\
&\medmath{=P\left(\left| \left(\beta R_{\textrm{s}} - \beta' R_{\textrm{s}}' \right) + \left(\gamma R_{\textrm{d}} - \gamma' R_{\textrm{d}}'\right) \right|<\zeta \mbox{ and } ||\alpha_{k}|^{2} - \eta(\beta' R_{\textrm{s}}' + \gamma' R_{\textrm{d}}') |<||\alpha_{k}|^{2} - \eta(\beta' R_{\textrm{s}} + \gamma' R_{\textrm{d}}) |\right)}\notag \\
&\medmath{\leq P\left(\left| \left(\beta R_{\textrm{s}} - \beta' R_{\textrm{s}}' \right) + \left(\gamma R_{\textrm{d}} - \gamma' R_{\textrm{d}}'\right) \right|<\zeta\right)\leq P\left(\left| \gamma R_{\textrm{d}} - \gamma' R_{\textrm{d}}' \right|<\zeta\right)},
\end{align}
it can be inferred that the error probability tends to be $0$ as $\zeta \rightarrow 0$. Hence, the proposed algorithm provides good clustering performance with high probability.

\subsubsection{Surface forming}

The purpose of surface forming is to derive the layouts and reflection properties of visible front surfaces, which jointly make up the whole ambient environment. Specifically, the layout refers to the line equation and endpoints of each surface, and can be determined through the coordinates of targets. The reflection property is simply the label of each cluster. We assume the ambient environment is composed of $L$ reflective surfaces which are directly visible by the AP and can be derived from the target clusters, and besides, there are $L_{\textrm{s}}$ supplementary surfaces which are directly invisible and are created when there are two orthogonally adjoining reflective surfaces. More details are provided below.

For clarity, let us focus on a specific cluster $\mathcal{C}_{l}=\{\textrm{T}_{k_{l,1}},\cdots,\textrm{T}_{k_{l,J_{l}}}\}$ that is labelled with $( \hat{R}_{\textrm{s},k_{l,1}},\hat{R}_{\textrm{d},k_{l,1}},\hat{\mathit{\Phi}}_{\textrm{re},k_{l,1}} )$. It is needed to determine a line segment in the 2-D plane that covers the coordinates $(x_{k_{l,1}},y_{k_{l,1}}),\cdots,(x_{k_{l,J_{l}}},y_{k_{l,J_{l}}})$, where $x_{k} \triangleq \textrm{c}\tau_{k}\cos(\theta_{k})$ and $y_{k} \triangleq \textrm{c}\tau_{k}\sin(\theta_{k})$. As the delay measurements are accurate, the line function can be calculated based on $(x_{k_{l,1}},y_{k_{l,1}})$ and $(x_{k_{l,J_{l}}},y_{k_{l,J_{l}}})$, thus inducing
\begin{equation}
y=a_{l}x+b_{l},
\label{line_point_slope}
\end{equation}
where
\begin{equation}
a_{l} = \frac{y_{k_{l,J_{l}}} - y_{k_{l,1}}}{x_{k_{l,J_{l}}} - x_{k_{l,1}}},~~b_{l} = -a_{l}x_{k_{l,1}} + y_{k_{l,1}}.
\end{equation}
Notably, when the delay measurements are impaired by ramp nonlinearity \cite{ayhan2016impact} or other factors, the least squares fitting technique should be utilized to obtain the best-fit line. Capitalizing on the results above, the $l$-th surface is characterized by linear equation (\ref{line_point_slope}), endpoints $(x_{k_{l,1}},y_{k_{l,1}})$, $(x_{k_{l,J_{l}}},y_{k_{l,J_{l}}})$ and reflection property $( \hat{R}_{\textrm{s},k_{l,1}},\hat{R}_{\textrm{d},k_{l,1}},\hat{\mathit{\Phi}}_{\textrm{re},k_{l,1}} )$.

Next, recall the rectangular object assumption, we see two adjacent surfaces might belong to the same rectangular reflector. Considering this, the $l$-th and the $(l+1)$-th\footnotemark[2] reflective surface are paired if two conditions are satisfied. First is the orthogonality, i.e., \footnotetext[2]{$l+1$ here means $\textrm{mod}({l,L})+1$.}
\begin{equation}
\medmath{\left|\left( x_{k_{l,J_{l}}} -  x_{k_{l,1}} \right)\left( x_{k_{l+1,J_{l+1}}} -  x_{k_{l+1,1}} \right) + \left( y_{k_{l,J_{l}}} -  y_{k_{l,1}} \right)\left( y_{k_{l+1,J_{l+1}}} -  y_{k_{l+1,1}} \right)\right|<\varepsilon},
\label{condition2}
\end{equation}
where $\varepsilon$ denotes a small positive quantity. The second condition excludes the wall surfaces via
\begin{equation}
x_{\textrm{ip},l}\left( x_{k_{l,J_{l}}} -  x_{k_{l,1}} \right)+y_{\textrm{ip},l}\left( y_{k_{l,J_{l}}} -  y_{k_{l,1}} \right)<0,
\label{condition1}
\end{equation}
where $(x_{\textrm{ip},l},y_{\textrm{ip},l})=(\frac{-b_{l}+b_{l+1}}{a_{l}-a_{l+1}},\frac{a_{l}b_{l+1}-a_{l+1}b_{l}}{a_{l}-a_{l+1}})$ is the intersection point of surface $l$ and $l+1$. This inequality guarantees that the included angle between vector $(x_{\textrm{ip},l},y_{\textrm{ip},l})$ and $(x_{k_{l,J_{l}}} -  x_{k_{l,1}},y_{k_{l,J_{l}}} -  y_{k_{l,1}})$ is an obtuse angle. When these two conditions are satisfied, we should replace $(x_{k_{l,J_{l}}},y_{k_{l,J_{l}}})$ and $(x_{k_{l+1,1}},y_{k_{l+1,1}})$ with $(x_{\textrm{ip},l},y_{\textrm{ip},l})$ in order to describe the corner of a rectangular object. Also, two \emph{supplementary} surfaces are created as follows:
\begin{equation}
y = a_{l}x+b'_{l}, ~\mbox{ and }~ y = a_{l+1}x+b'_{l+1},
\end{equation}
where $b'_{l}=-a_{l}x_{k_{l+1,J_{l+1}}} + y_{k_{l+1,J_{l+1}}}$, $b'_{l+1}=-a_{l+1}x_{k_{l,1}} + y_{k_{l,1}}$. The endpoints of these two surfaces are $(x_{k_{l,1}},y_{k_{l,1}})$, $(x'_{\textrm{ip},l},y'_{\textrm{ip},l})$, and $(x'_{\textrm{ip},l},y'_{\textrm{ip},l})$, $(x_{k_{l+1,J_{l+1}}},y_{k_{l+1,J_{l+1}}})$. Particularly, $(x'_{\textrm{ip},l},y'_{\textrm{ip},l})=(\frac{-b'_{l}+b'_{l+1}}{a_{l}-a_{l+1}},\frac{a_{l}b'_{l+1}-a_{l+1}b'_{l}}{a_{l}-a_{l+1}})$ represents the intersection point of supplementary surfaces. Repeating this process, $L_{\textrm{s}}$ supplementary surfaces are formed.


\subsection{AP-CR: Ambient Perception based Channel Reconstruction}

Given the ambient perceptions, once the AP is informed of UEs' positions, it is able to enumerate all possible propagation paths towards them through ray-tracing technique.

To proceed, suppose there exist $M$ UEs in the room. The $m$-th UE comprising a uniform linear array (ULA) with $N_{\textrm{ue},m}$ antennas is located at $\left(x_{\textrm{ue},m},y_{\textrm{ue},m}\right)$. UEs' positions are estimated using advanced positioning techniques (e.g., \cite{lin20183,jia2018motion}) as
\begin{equation}
\left(\hat{x}_{\textrm{ue},m},\hat{y}_{\textrm{ue},m}\right) = \left(x_{\textrm{ue},m},y_{\textrm{ue},m}\right) + \left(\Delta x_{m},\Delta y_{m}\right),
\end{equation}
where $\Delta x_{m}\sim \mathcal{N}\left(0,\sigma^{2}_{\textrm{e}}\right)$ and $\Delta y_{m}\sim \mathcal{N}\left(0,\sigma^{2}_{\textrm{e}}\right)$ denote the positioning error, $\sigma^{2}_{\textrm{e}}$ is the variance. Intuitively, better positioning techniques give rise to smaller $\sigma^{2}_{\textrm{e}}$. Nonetheless, details are out the scope of this paper.

Next, leveraging the fact that the aperture of an isotropic antenna satisfies $A_{\textrm{iso}}\propto \lambda^{2}$ in 3-D space, it is practical to set the 2-D aperture of such antenna as $\lambda$. In this sense, the aperture of the $m$-th UE becomes $A_{\textrm{ue},m}=N_{\textrm{ue},m}\lambda$ if antennas are separated by $\lambda$. The orientation of the ULA is $\varphi_{\textrm{ue},m}$\footnotemark[3]. \footnotetext[3]{In practice, the array orientation $\varphi_{\textrm{ue}}$ needs to be estimated. However, we assume it is known in this paper and leave its estimation for our future work.} Then, channel reconstruction can be realized via two steps, namely,
\begin{itemize}
    \item LoS path detection. The AP determines the existence, AoD and AoA of the LoS path.
    \item NLoS path computation. The AP computes how to illuminate the UE with NLoS specular reflections and derives the corresponding AoD/AoA pairs.
\end{itemize}
In the first step, to judge the existence of the LoS path, we define
\begin{equation}
\Theta\left(x,y\right)\triangleq\left\{\begin{matrix}
\textrm{mod}(\arctan(\frac{y}{x}),2\pi),\quad~~~~~x> 0~
\\
\arctan(\frac{y}{x}) + \pi,~~~~~~~~~~~~~~x<0~
\\
\pi - \frac{y}{2|y|}\pi,~~~~~~~~~~~~~~~~~~~~x = 0.
\end{matrix}\right.
\label{angular_function}
\end{equation}
As such, the $l$-th reflective surface induces an angular range
\begin{equation}
\Omega_{l}=\left\{\begin{matrix}
[\Theta(x_{k_{l,1}},y_{k_{l,1}}),\Theta(x_{k_{l,J_{l}}},y_{k_{l,J_{l}}})),~~~~~~~~~~~~~~\Theta(x_{k_{l,1}},y_{k_{l,1}})<\Theta(x_{k_{l,J_{l}}},y_{k_{l,J_{l}}})~
\\
[\Theta(x_{k_{l,1}},y_{k_{l,1}}),2\pi)\cup[0,\Theta(x_{k_{l,J_{l}}},y_{k_{l,J_{l}}})),~~~\Theta(x_{k_{l,1}},y_{k_{l,1}})>\Theta(x_{k_{l,J_{l}}},y_{k_{l,J_{l}}}).
\end{matrix}\right.
\label{angular_range}
\end{equation}
Then, the existence of the LoS path is determined by the following proposition.

\emph{Proposition 2:} The existence of the LoS path can be denoted symbolically as
\begin{align}
I_{m}^{(0)}
=\left(\nexists l\in \left\{ 1,\cdots,L \right\},~\Theta(\hat{x}_{\textrm{ue},m},\hat{y}_{\textrm{ue},m})\in \Omega_{l}
\mbox{~and~} \left|\frac{b_{l}}{y_{\textrm{ue},m} - a_{l}x_{\textrm{ue},m}}\right|<1\right).
\label{LOS_indicator_bu}
\end{align}

\begin{IEEEproof}
See Appendix \ref{Proof_P2}.
\end{IEEEproof}

Provided that $I_{m}^{(0)}=1$, the AoD and AoA are easily obtained by
\begin{align}
&\theta_{\textrm{AoD},m}^{(0)} = \Theta(\hat{x}_{\textrm{ue},m},\hat{y}_{\textrm{ue},m}), \label{LOS_AoD} \\
&\theta_{\textrm{AoA},m}^{(0)} = \textrm{mod}(\Theta(\hat{x}_{\textrm{ue},m},\hat{y}_{\textrm{ue},m}) - \varphi_{\textrm{ue},m} + \pi,2\pi). \label{LOS_AoA}
\end{align}
Depending on $\theta_{\textrm{AoA},m}^{(0)}$, the effective UE aperture is $A_{\textrm{ue},m}^{(0)} = \lambda + \left(N_{\textrm{ue},m}-1\right)\lambda \left|\sin(\theta_{\textrm{AoA},m}^{(0)})\right|$. Hence, the propagation gain of the LoS path is given by
\begin{equation}
\alpha_{m}^{(0)}=e^{j\frac{2\pi D_{m}^{(0)}}{\lambda}}\sqrt{\min\left(\frac{A_{\textrm{ue},m}^{(0)}}{W(\theta_{\textrm{AoD},m}^{(0)},N_{\textrm{T}}) D_{m}^{(0)}},1\right)},
\label{propagation_gain_LOS}
\end{equation}
where $D_{m}^{(0)}=\sqrt{x_{\textrm{ue},m}^{2}+y_{\textrm{ue},m}^{2}}$ is the distance from UE $m$ to the AP.

\begin{figure}[!t]\centering
\includegraphics[angle=0,scale=0.26]{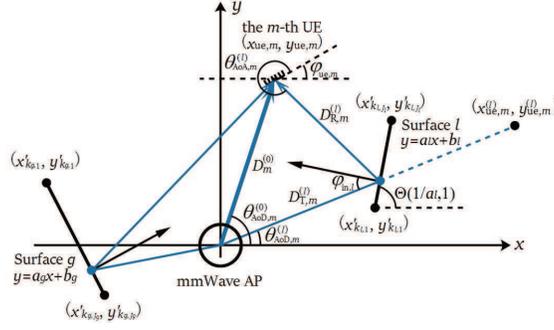}
\vspace{-0.5cm}
\caption{Illustration of NLoS path computation.}
\label{Fig_NLOS_computation}
\vspace{-0.5cm}
\end{figure}

Next, it is needed to determine the NLoS counterpart. For this purpose, let us look at the $l$-th reflective surface illustrated in Fig. \ref{Fig_NLOS_computation}. The mirror image of $\left(x_{\textrm{ue},m},y_{\textrm{ue},m}\right)$ with respect to it has coordinate $(x_{\textrm{ue},m}^{(l)},y_{\textrm{ue},m}^{(l)})$, in which
\begin{equation}
x_{\textrm{ue},m}^{(l)}=\frac{(1-a_{l}^{2})\hat{x}_{\textrm{ue},m}-2a_{l}b_{l}+2a_{l}\hat{y}_{\textrm{ue},m}}{a_{l}^{2}+1},~~ y_{\textrm{ue},m}^{(l)}=\frac{2a_{l}\hat{x}_{\textrm{ue},m}+2b_{l}-(1-a_{l}^{2})\hat{y}_{\textrm{ue},m}}{a_{l}^{2}+1}.
\end{equation}
For higher propagation gain, the NLoS path had better involve specular reflection. Thus, the coordinate of the reflection point on the $l$-th reflective surface is calculated as
\begin{equation}
\left(x_{\textrm{re},m}^{(l)},y_{\textrm{re},m}^{(l)}\right) = (\frac{b_{l}x_{\textrm{ue},m}^{(l)}}{y_{\textrm{ue},m}^{(l)} -a_{l}x_{\textrm{ue},m}^{(l)}},\frac{b_{l}y_{\textrm{ue},m}^{(l)}}{y_{\textrm{ue},m}^{(l)} -a_{l}x_{\textrm{ue},m}^{(l)}}).
\end{equation}
On this basis, the following proposition is derived.

\emph{Proposition 3:} The existence of the $l$-th NLoS path can be represented by
\begin{align}
\medmath{I_{m}^{(l)}}
&\medmath{=\left(\Theta(x_{\textrm{ue},m}^{(l)},y_{\textrm{ue},m}^{(l)}) \in \Omega_{l} \mbox{~and~} \left|\frac{b_{l}}{y_{\textrm{ue},m}^{(l)} -a_{l}x_{\textrm{ue},m}^{(l)}}\right|<1\right) \cap \Big( \nexists n\in \left\{ 1,\cdots,L+L_{\textrm{s}} \right\}\backslash \{l\},} \notag\\
&\medmath{\Theta\Big(\frac{\hat{x}_{\textrm{ue},m}y_{\textrm{re},m}^{(l)} - x_{\textrm{re},m}^{(l)}\hat{y}_{\textrm{ue},m} + b_{n}(x_{\textrm{re},m}^{(l)} - \hat{x}_{\textrm{ue},m})}{(y_{\textrm{re},m}^{(l)} - \hat{y}_{\textrm{ue},m})-a_{n}(x_{\textrm{re},m}^{(l)} - \hat{x}_{\textrm{ue},m})}, \frac{a_{n}(\hat{x}_{\textrm{ue},m}y_{\textrm{re},m}^{(l)} - x_{\textrm{re},m}^{(l)}\hat{y}_{\textrm{ue},m}) + b_{n}(y_{\textrm{re},m}^{(l)} - \hat{y}_{\textrm{ue},m})}{(y_{\textrm{re},m}^{(l)} - \hat{y}_{\textrm{ue},m})-a_{n}(x_{\textrm{re},m}^{(l)} - \hat{x}_{\textrm{ue},m})} \Big)} \notag \\
&\medmath{\in \Omega_{n} \mbox{~and~} \left|\frac{ a_{n}x_{\textrm{re},m}^{(l)} + b_{n} - y_{\textrm{re},m}^{(l)} }{ (y_{\textrm{re},m}^{(l)}-\hat{y}_{\textrm{ue},m}) - a_{n}(x_{\textrm{re},m}^{(l)}-\hat{x}_{\textrm{ue},m}) }\right|<1 \mbox{~and~} \left|\frac{ a_{n}\hat{x}_{\textrm{ue},m} + b_{n} - \hat{y}_{\textrm{ue},m} }{ (y_{\textrm{re},m}^{(l)}-\hat{y}_{\textrm{ue},m}) - a_{n}(x_{\textrm{re},m}^{(l)}-\hat{x}_{\textrm{ue},m}) }\right|<1\Big)}.
\label{NLOS_indicator}
\end{align}

\begin{IEEEproof}
See Appendix \ref{Proof_P3}.
\end{IEEEproof}

Similarly, if $I_{m}^{(l)}=1$, the AoD and AoA of the corresponding path are given by
\begin{align}
&\theta_{\textrm{AoD},m}^{(l)} = \Theta(x_{\textrm{ue},m}^{(l)},y_{\textrm{ue},m}^{(l)}), \label{AoD_l}  \\
&\theta_{\textrm{AoA},m}^{(l)} = \textrm{mod}(2 \Theta(1/a_{l},1) - \Theta(x_{\textrm{ue},m}^{(l)},y_{\textrm{ue},m}^{(l)}) - \varphi_{\textrm{ue},m} + \pi,2\pi). \label{AoA_l}
\end{align}
Invoking (\ref{reflection_coefficient}), the reflection coefficient of the $l$-th propagation path can be expressed as
\begin{align}
&\medmath{\alpha_{m}^{(l)} = e^{-j(\hat{\mathit{\Phi}}_{\textrm{re},k_{l,1}}-\frac{2\pi (D_{\textrm{T},m}^{(l)}+D_{\textrm{R},m}^{(l)})}{\lambda})}\times} \notag\\
&\medmath{\sqrt{\eta\big( \min\big(\frac{A_{\textrm{ue},m}^{(l)}}{W(\theta_{\textrm{AoD},m}^{(l)},N_{\textrm{T}})(D_{\textrm{T},m}^{(l)}+D_{\textrm{R},m}^{(l)})},1\big)\hat{R}_{\textrm{s},k_{l,1}} +  \frac{\sin^{2}(\Theta(\frac{1}{a_{l}},1) - \Theta(x_{\textrm{ue},m}^{(l)},y_{\textrm{ue},m}^{(l)}))A_{\textrm{ue},m}^{(l)}}{\sqrt{4 (D_{\textrm{R},m}^{(l)})^{2} + (A_{\textrm{ue},m}^{(l)})^{2}}}\hat{R}_{\textrm{d},k_{l,1}} \big)}},
\label{reflection_coefficient_l}
\end{align}
where $A_{\textrm{ue},m}^{(l)} = \lambda + (N_{\textrm{ue},m}-1)\lambda|\sin(\theta_{\textrm{AoA},m}^{(l)})|$ is the effective aperture, $D_{\textrm{T},m}^{(l)} =\sqrt{( x_{\textrm{re},m}^{(l)})^{2} + ( y_{\textrm{re},m}^{(l)})^{2}}$ and $D_{\textrm{R},m}^{(l)} =\sqrt{( x_{\textrm{re},m}^{(l)} - x_{\textrm{ue},m})^{2} + ( y_{\textrm{re},m}^{(l)} - y_{\textrm{ue},m})^{2}}$ are the distances from the reflection point to the AP and the $m$-th UE, respectively.

After LoS detection and NLoS computation, the mmWave channel between the AP and the $m$-th UE can be reconstructed as
\begin{equation}
\hat{\textbf{H}}_{m} = \sum_{l=0}^{L}I_{m}^{(l)} \alpha_{m}^{(l)}\textbf{a}_{\textrm{ue}}\left(\theta_{\textrm{AoA},m}^{(l)},N_{\textrm{ue},m}\right)\textbf{a}_{\textrm{T}}^{H}\left(\theta_{\textrm{AoD},m}^{(l)},N_{\textrm{T}}\right),
\label{channel_model}
\end{equation}
where $\textbf{a}_{\textrm{ue}}(\theta,N_{\textrm{ue},m})
=[1,e^{-j2\pi \cos (\theta)},\dots,e^{-j2\pi(N_{\textrm{ue},m}-1)\cos (\theta)}]^{T}$ is the receive steering vector of the $m$-th UE.

\subsection{Discussions on the Reconstruction Performance}\label{discussion}

Given the reconstructed channel, let us look into the factors influencing the reconstruction performance. Intuitively, the reconstructed channel should be ideal if the ambient environment is perfectly perceived. However, this is extremely difficult. Additionally, positioning error incurs miscalculations of reflection paths, thus producing erroneous channel descriptions.

Further, let us investigate the ambient perception imperfectness. In the proposed scheme, the reason for such imperfectness is mainly threefold: a) insufficient antennas; b) insufficient probing beams; c) inadequate perception methods. Regarding the number of antennas, it can be inferred from \eqref{bwidth} and \eqref{steering_vector_n} that more antennas lead to narrower beams and smaller illumination regions. Recall that the echoes reflected by multiple objects are discarded in the sensing procedure, a smaller illumination region reduces the probability of ineffective sensing results. Consequently, we conclude that massive antennas improve the performance of ambient perception. Secondly, a small number of probing beams might be too sparse for the AP to perceive the ambient reflectors. Recall the endpoints of the perceived reflective surfaces, such sparsity inevitably makes the perceived surfaces (i.e., 2-D segments) much \emph{shorter} than the actual ones. Incurred by this, the AP cannot judge the existences of the LoS and NLoS path accurately. Accordingly, the proposed scheme prefers more probing beams. Thirdly, the proposed ambient perception scheme is rather simple. The AP is able to perceive the surfaces in the LoS area and a small number of supplementary surfaces in the shadow area, yet being incapable of handling every surface in the room. In fact, there might exist surfaces that block the NLoS paths and cannot be perceived \eqref{NLOS_indicator}. To cope with this problem, a possible solution based on UEs' feedbacks is presented in the next subsection.

Another important factor lies in the positioning error. Considering the AoD/AoA pairs of the LoS and NLoS paths, i.e., \eqref{LOS_AoD}, \eqref{LOS_AoA}, \eqref{AoD_l} and \eqref{AoA_l}, all of them are impacted by the positioning error. In case of large deviations, the transmit and receive beams are probably no longer aligned with the propagation channel, making the actual path gains much smaller than the reconstructed gains \eqref{propagation_gain_LOS} and \eqref{reflection_coefficient_l}. To mitigate this effect, an intuitive solution is to apply beamformers with larger beamwidth in subsequent communications, at the price of lower beamforming gain and higher mutual interference. Note that the beamwidth no longer conforms to \eqref{bwidth}, the path gains and the induced channel description should be modified. Apparently, this yields a beamwidth optimization problem when the positioning error level is acknowledged. Interested readers are referred to \cite{wildman2014on} and the references therein.

\subsection{AP-CRF: Ambient Perception based Channel Reconstruction with Feedback}

Similar to the codebook-based channel estimation scheme \cite{alkhateeb2014channel}, we exploit the fact that UEs are able to inform the AP of the possible AoAs. On this basis, the $m$-th UE examines whether data signals can be received from direction $\theta_{\textrm{AoA},m}^{(l)}$. When only interfering signals are detected, the AP infers that the $l$-th path in $\hat{\textbf{H}}_{m}$ is blocked by some unknown surfaces, which induces a negative feedback $F_{m}^{(l)}=0$. Otherwise, $F_{m}^{(l)}=1$ is sent back to the AP.

Upon receiving the feedback $\{F_{m}^{(l)}\}_{l=0,\dots,L}$, the AP recalculates the mmWave channel as
\begin{equation}
\hat{\textbf{H}}_{m} = \sum_{l=0}^{L}F_{m}^{(l)}\alpha_{m}^{(l)}\textbf{a}_{\textrm{ue}}\left(\theta_{\textrm{AoA},m}^{(l)},N_{\textrm{ue},m}\right)\textbf{a}_{\textrm{T}}^{H}\left(\theta_{\textrm{AoD},m}^{(l)},N_{\textrm{T}}\right).
\label{channel_model_pro}
\end{equation}
Based on such channel, accurate beamformers can be designed. In case of $\hat{\textbf{H}}_{m}=\textbf{0}$, i.e., there is neither LoS nor NLoS path, the AP has to communicate with the $m$-th UE using sub-6GHz band systems.

\section{MmWave Communication}

Due to the sparse characteristics, the mmWave channel is probably composed of several paths with clearly distinguishable gains. As such, it is practical to restrict our attention to single-beam beamforming along the strongest path.

In particular, we consider the scenario that the AP transmits one data stream to each UE with analog beamforming. The strongest path is undoubtedly the LoS path if it exists. Once the LoS path is blocked, the strongest path refers to the NLoS path with the highest reflection coefficient. Taking the $m$-th UE for instance, the index of the strongest path is derived as
\begin{equation}
l_{\textrm{str}} = \underset{l\in \{0,1,\cdots,L\}}{\arg \max}|F_{m}^{(l)}\alpha_{m}^{(l)}|.
\end{equation}
If all the paths are blocked, then no data is transmitted to that UE. In this setting, the received data signal at the $m$-th UE can be modeled by
\begin{equation}
\textbf{y}_{m} = \textbf{H}_{m}\sum_{i=1}^{M}\textbf{x}_{i} + \textbf{n}_{m},
\label{receive_data_signal}
\end{equation}
where $\textbf{H}_{m}\in \mathbb{C}^{N_{\textrm{ue},m}\times N_{\textrm{T}}}$ denotes the channel from the AP to the $m$-th UE, $\textbf{x}_{i}= \sqrt{P_{\textrm{T}}} F_{i}^{(l_{\textrm{str}})}\textbf{f}_{\textrm{RF},i}^{(l_{\textrm{str}})} s_{i}$ is the data vector sent to the $i$-th UE, $\textbf{f}_{\textrm{RF},i}^{(l_{\textrm{str}})}\in \mathbb{C}^{N_{\textrm{T}}\times 1}$ is a analog transmit beamformer, $s_{i}\sim \mathcal{N}\left(0,1\right)$ is the transmit data signal, $\textbf{n}_{m}\sim \mathcal{CN}\left(\textbf{0},\sigma_{\textrm{n}}^{2}\textbf{I}_{N_{\textrm{ue},m}}\right)$ is the complex Gaussian noise.

After applying the receive beamformer $\textbf{w}_{\textrm{RF},m}^{(l_{\textrm{str}})}\in \mathbb{C}^{1\times N_{\textrm{ue},m}}$, the post-processed signal becomes
\begin{equation}
z_{m} = \sqrt{P_{\textrm{T}}}\textbf{w}_{\textrm{RF},m}^{(l_{\textrm{str}})}\textbf{H}_{m}\sum_{i=1}^{M}F_{i}^{(l_{\textrm{str}})}\textbf{f}_{\textrm{RF},i}^{(l_{\textrm{str}})} s_{i} + \bar{n}_{m},
\label{receive_data_signal_pro}
\end{equation}
where $\bar{n}_{m}\triangleq\textbf{w}_{\textrm{RF},m}^{(l_{\textrm{str}})}\textbf{n}_{m}\sim \mathcal{CN}\left(0,\sigma_{\textrm{n}}^{2}\right)$ is the equivalent noise.

From the reconstructed channel (\ref{channel_model_pro}), the optimal transmit and receive beamformers are given by
\begin{equation}
\textbf{f}_{\textrm{RF},m}^{(l_{\textrm{str}})} = \frac{\textbf{a}_{\textrm{T}}\left(\theta_{\textrm{AoD},m}^{(l_{\textrm{str}})},N_{\textrm{T}}\right)}{\sqrt{N_{\textrm{T}}}}, ~\mbox{ and }~
\textbf{w}_{\textrm{RF},m}^{(l_{\textrm{str}})}=\frac{\textbf{a}_{\textrm{R}}^{H}\left(\theta_{\textrm{AoA},m}^{(l_{\textrm{str}})},N_{\textrm{ue},m}\right)}{\sqrt{N_{\textrm{ue},m}}}.
\label{single_beam_beamformer}
\end{equation}
Substituting (\ref{single_beam_beamformer}) into (\ref{receive_data_signal_pro}), if there exists no positioning error, i.e., $\sigma_{\textrm{e}}^{2}=0$, then
\begin{equation}
z_{m} \approx \mbox{$F_{m}^{(l_{\textrm{str}})}\alpha_{m}^{(l_{\textrm{str}})}\sqrt{ P_{\textrm{T}}N_{\textrm{T}}N_{\textrm{ue},m}} s_{m} + u_{m} + \bar{n}_{m}$},
\label{receive_single}
\end{equation}
where $u_{m}$ represents the interference term and is given by
\begin{align}
u_{m} = \sqrt{P_{\textrm{T}}}\textbf{w}_{\textrm{RF},m}^{(l_{\textrm{str}})}\textbf{H}_{m}\sum_{i\neq m}F_{i}^{(l_{\textrm{str}})}\textbf{f}_{\textrm{RF},i}^{(l_{\textrm{str}})} s_{i}.
\label{interference}
\end{align}
Based on (\ref{receive_single}), the transmission rate from the AP to the $m$-th UE can be approximated by
\begin{equation}
R_{m}\approx \textrm{B}\times\mathbb{E}\left[\log_{2}\left(1+\frac{(F_{m}^{(l_{\textrm{str}})}\alpha_{m}^{(l_{\textrm{str}})})^{2} P_{\textrm{T}}N_{\textrm{T}}N_{\textrm{ue},m}\left|s_{m}\right|^{2}}{ \left|u_{m}\right|^{2} + \sigma_{\textrm{n}}^{2}}\right)\right],
\label{rate_single}
\end{equation}
where $\textrm{B}$ denotes the bandwidth.

\section{Numerical Results}

In this section, we evaluate the proposed system with Monte Carlo simulations. Simulation parameters are shown in TABLE IV. For each simulation, a random indoor scenario is constructed with the model in Section \ref{indoor_scenario}. Nonetheless, these scenarios share the same table of reflection property (see TABLE V), in which the reflectance and phase shift are determined using $\textrm{U}\left(0,1\right)$ and $\textrm{U}\left(0,2\pi\right)$, respectively. The first five properties are randomly assigned to the interior objects. The last one corresponds to the wall.

\begin{table}[!t]
\renewcommand\arraystretch{0.6}
\begin{center}
\caption[]{Simulation Parameters}
\begin{tabular}{|c|c|}
\hline
Length of the room, $L_{\textrm{r}}$                         &   $20$m\\ \hline
Width of the room, $W_{\textrm{r}}$                          &   $15$m\\ \hline
Position of the AP                                           &   $12.5$m from the left, $5$m from the bottom\\ \hline
Parameter of poisson point process, $\lambda_{\textrm{ppp}}$ &   $0.015$\\ \hline
Maximum length of the object, $L_{\textrm{o}}$               &   $5$m\\ \hline
Maximum width of the object, $W_{\textrm{o}}$                &   $5$m\\ \hline
Operating frequency of mmWave, $f_{\textrm{c}}$              &   $60$GHz  \\ \hline
Wavelength of mmWave, $\lambda=\frac{\textrm{c}}{f_{\textrm{c}}}$    &   $0.005$m \\ \hline
Bandwidth of mmWave, $B$                                     &   $1$GHz  \\ \hline
Number of probing beams, $K$                                 &   $256,512,1024$ \\ \hline
Number of AP's antennas, $N_{\textrm{T}}=N_{\textrm{R}}$     &   $64,96,\cdots,256$  \\ \hline
Width of AP's transmit beams, $W$                            &   Calculated using (\ref{bwidth}) \\ \hline
AP's array aperture, $A=2r$                                  &   $\frac{N_{\textrm{T}}\lambda}{\pi}$  \\ \hline
Table of reflection property, $\{( R_{\textrm{s}},R_{\textrm{d}}, \mathit{\Phi}_{\textrm{re}})\}$& See TABLE V  \\ \hline
Inherent reflected ratio, $\eta$                             &   $0.25$  \\ \hline
Miscalculation threshold, $\zeta$                            &   $5\times 10^{-7}$  \\ \hline
Number of UEs, $M$                                           &   $10$ \\ \hline
Number of UE's antennas, $N_{\textrm{ue},m}$                 &   $32$ \\ \hline
UE's array aperture, $A_{\textrm{ue},m}=N_{\textrm{ue},m}\lambda$   &   $0.16$m \\ \hline
Variance of positioning error, $\sigma^{2}_{\textrm{e}}$     &   $0,0.0004,0.0008\cdots,0.004$ \\ \hline
Array orientation of UE, $\varphi_{\textrm{ue},m}$           &   $\textrm{U}\left(0,\pi\right)$ \\ \hline
Normalized transmit power of each UE, $P_{\textrm{T}}$       &   $1$ \\ \hline
\end{tabular}
\end{center}
\vspace{-0.75cm}
\end{table}

\begin{table}[!t]
\renewcommand\arraystretch{0.6}
\begin{center}
\caption[]{Table of reflection property}
\begin{tabular}{|c|c|c|c|}
\hline
Index       &$R_{\textrm{s}}\sim \textrm{U}\left(0,1\right)$       &$R_{\textrm{d}}\sim \textrm{U}\left(0,1\right)$       &$\mathit{\Phi}_{\textrm{re}}\sim \textrm{U}\left(0,2\pi\right)$               \\ \hline
1           &0.6606                     &0.6781                 &$2.6134$    \\ \hline
2           &0.3286                     &0.6464                 &$2.2530$    \\ \hline
3           &0.5233                     &0.9288                 &$1.1897$    \\ \hline
4           &0.2865                     &0.2412                 &$4.3408$    \\ \hline
5           &0.8745                     &0.3547                 &$4.7810$    \\ \hline
6           &0.5736                     &0.3262                 &$2.4454$    \\ \hline
\end{tabular}
\end{center}
\vspace{-1.5cm}
\end{table}

\begin{figure}[!t]\centering
\includegraphics[angle=0,scale=0.57]{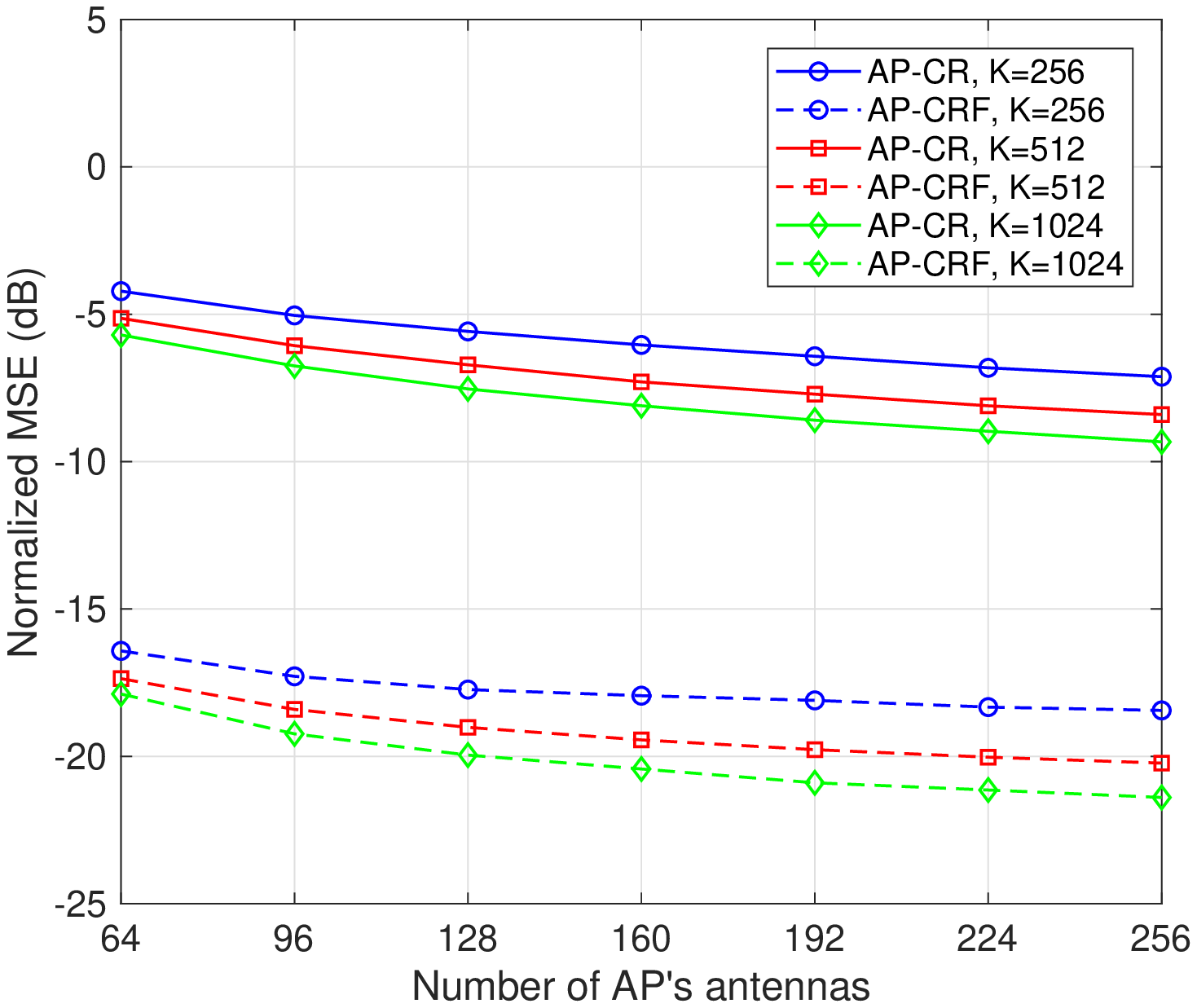}
\vspace{-0.5cm}
\caption{Normalized channel reconstruction MSE for AP-CR and AP-CRF, $\sigma^{2}_{\textrm{e}}=0$.}
\label{Fig_Channel_Estimation_MSE}
\vspace{-0.75cm}
\end{figure}

In the first example, we examine the channel reconstruction performance from the perspective of perception imperfectness. Positioning error is not considered, i.e., $\sigma^{2}_{\textrm{e}}=0$. The channel produced by ray-tracing with perfect ambient information is treated as the benchmark $\textbf{H}_{m}$. Based on this, the normalized mean square error (NMSE), i.e., $\sum_{m=1}^{M}\|\hat{\textbf{H}}_{m}-\textbf{H}_{m}\|^{2}/\|\textbf{H}_{m}\|^{2}$, is shown in Fig. \ref{Fig_Channel_Estimation_MSE}. The special case $\textbf{H}_{m}=0$ has been discarded. As expected, the MSE curves decrease with the number of transmit antennas and the number of probing beams. It can be observed that, in high $N_{\textrm{T}}$ regime, increasing the probing beams provides a larger gain. This is because echo signals are more informative when probing beams are narrow. By comparing the MSEs of AP-CR and AP-CRF, it is verified that UEs' feedbacks contribute to better channel estimates. The huge gap between them indicates that the bottleneck of AP-CR lies in the perception method. To further reduce the reconstruction erro, improved ambient perception schemes are urgently needed.

\begin{figure}[!t]\centering
\includegraphics[angle=0,scale=0.57]{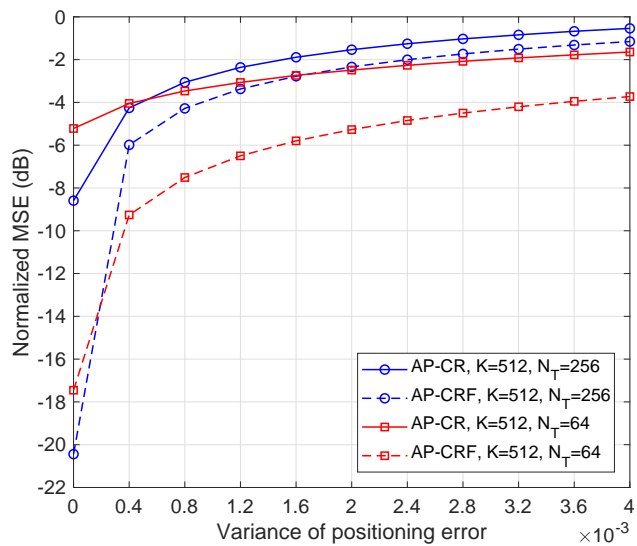}
\vspace{-0.5cm}
\caption{Normalized channel reconstruction MSE for AP-CR and AP-CRF, $K=512$, $N_{\textrm{T}}=64,256$.}
\label{Fig_Channel_Estimation_MSE_PE}
\vspace{-0.5cm}
\end{figure}

Next, Fig. \ref{Fig_Channel_Estimation_MSE_PE} investigates the reconstruction NMSE incurred by positioning error. For ease of comparison, $K=512$ probing beams and $N_{\textrm{T}}=64,~256$ transmit antennas are considered. In this figure, the MSE curves increase with the positioning error. The increasing rate is faster when more antennas are exploited, which verifies the fact that narrow beams are more sensitive to positioning error. More specifically, for AP-CR with $\sigma^{2}_{\textrm{e}}>0.0004$ and AP-CRF with $\sigma^{2}_{\textrm{e}}>0.0002$, $N_{\textrm{T}}=64$ antennas (i.e., wider beams) provide better reconstruction performance than the $N_{\textrm{T}}=256$ test case. These observations verify the discussions in Section \ref{discussion}.

\begin{figure}[!t]\centering
\includegraphics[angle=0,scale=0.57]{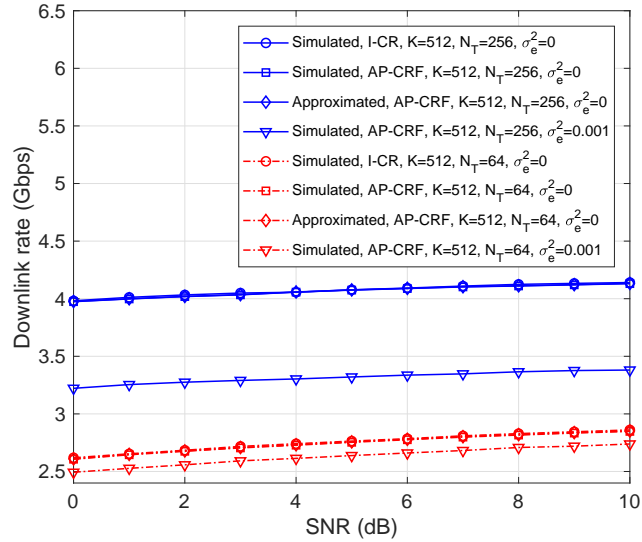}
\vspace{-0.5cm}
\caption{Downlink rate with single-beam beamforming, $K=512$, $N_{\textrm{T}}=64,~256$, $\sigma_{\textrm{e}}^{2}=0,~0.001$.}
\label{Fig_Downlink_rate_single}
\vspace{-0.5cm}
\end{figure}

Fig. \ref{Fig_Downlink_rate_single} depicts the downlink rates with single-beam beamforming using $K=512$ probing beams and $N_{\textrm{T}}=64,~256$ antennas. The variance of positioning error is selected as $\sigma_{\textrm{e}}^{2}=0,~0.001$. The curves produced by ideal channel reconstruction (I-CR) and $\sigma_{\textrm{e}}^{2}=0$ act as the benchmark. It is trivial that the downlink rate grows with the transmit signal-to-noise ratio (SNR) $P_{\textrm{T}}/\sigma_{\textrm{n}}^{2}$. When there exists no positioning error, the simulated curves of I-CR and AP-CRF are quite close, which validates the good performance of the proposed scheme. The approximated curve of AP-CRF is very accurate, thus verifying the effectiveness of \eqref{rate_single}. When $\sigma_{\textrm{e}}^{2}$ increases to $0.001$, rate degradations are observed. This is intuitive from the perspective of channel reconstruction MSE. In particular, the $N_{\textrm{T}}=256$ test case suffers about $0.8$Gbps ($\approx 20\%$) downlink rate decrease, while $N_{\textrm{T}}=64$ reduces about $0.1$Gbps ($\approx 4\%$). This insightful observation coincides with the results in Fig. \ref{Fig_Channel_Estimation_MSE_PE} and verifies our analyses in Section III-C. The robustness of $N_{\textrm{T}}=64$ comes from the large illumination region of wide transmit beams. Furthermore, we see the overall performance of $N_{\textrm{T}}=256$ is still better in the current settings. When $\sigma_{\textrm{e}}^{2}$ exceeds a threshold, a small number of antennas are possible to outperform massive antennas. However, this is out the scope of this paper.

\section{Conclusion and Future Work}

In this paper, we have presented a mmWave joint radar and communication system. Invoking the indoor environment perceived by mmWave radar, accurate AoDs and AoAs from the mmWave AP to UEs are obtained. Based on a newly proposed mmWave reflection framework, the gain of each path is also successfully modeled. These results help the AP obtain high-accuracy channel estimates with low spectrum overhead. Even if UEs move to new positions, the perceived environment information makes it quite easy to re-calculate the propagation channel. Moreover, with the estimated CSI above, single-beam communication is investigated. Numerical results validate the effectiveness and good performance of the proposed ambient perception based schemes. The discussions on the robustness are also verified. In our future work, we plan to design a more robust perception algorithm and validate the ray-tracing framework with mmWave channel datasets as well as measurement studies.

\appendix

\subsection{Related functions in Algorithm 1}\label{Functions}

REFLECTANCE calculates the reflectance at a specific target. FLAG determines whether a target should be clustered. When two adjacent targets share true flags, CLUSTER expands or creates a cluster. Finally, TAIL2HEAD deals with the last target $T_{K}$ and the first one $T_{1}$.

\begin{table}[!ht]
\begin{algorithmic}[1]
\Function{REFLECTANCE}{$k,\tau,\varphi_{\textrm{in}}$}
\State Calculate $\beta$, $\gamma$ based on (\ref{beta_gamma_k});
\State For all reflectance pairs in $\{(R_{\textrm{s}},R_{\textrm{d}},\mathit{\Phi}_{\textrm{re}})\}$, find $( \hat{R}_{\textrm{s}},\hat{R}_{\textrm{d}} )=\underset{(R_{\textrm{s}},R_{\textrm{d}})}{\arg \min}| |\alpha|^{2} - \eta(\beta R_{\textrm{s}} + \gamma R_{\textrm{d}}) |$, the results are discarded if the minimum value exceeds $\zeta$;
\State \Return $( \hat{R}_{\textrm{s}},\hat{R}_{\textrm{d}} )$.
\EndFunction
\end{algorithmic}
\end{table}

\begin{table}[!ht]
\begin{algorithmic}[1]
\Function{FLAG}{$( \hat{R}_{\textrm{s},k},\hat{R}_{\textrm{d},k} ),( \hat{R}_{\textrm{s},k+1},\hat{R}_{\textrm{d},k+1} )$}
\If {$( \hat{R}_{\textrm{s},k},\hat{R}_{\textrm{d},k} )$ = $( \hat{R}_{\textrm{s},k+1},\hat{R}_{\textrm{d},k+1} )$ or $( \hat{R}_{\textrm{s},k},\hat{R}_{\textrm{d},k} )$ = REFLECTANCE$(k+1,\tau_{k+1},\vartheta_{k} + \Delta \theta - \pi/2)$}
    \State \Return $1$.
\EndIf
\EndFunction
\end{algorithmic}
\end{table}

\begin{table}[!ht]
\begin{algorithmic}[1]
\Function{CLUSTER}{$l,f_{k},f_{k+1},( \hat{R}_{\textrm{s},k},\hat{R}_{\textrm{d},k} )$}
\If {$f_{k+1}=1$}
    \State Put $\textrm{T}_{k+1}$ into the cluster that contains $\textrm{T}_{k}$ when $f_{k}=1$; Set $l=l+1$, $f_{k}=1$, create the $l$-th cluster as $\{\textrm{T}_{k},\textrm{T}_{k+1}\}$, and label it with reflectance $( \hat{R}_{\textrm{s},k},\hat{R}_{\textrm{d},k},\hat{\mathit{\Phi}}_{\textrm{re},k} )$ when $f_{k}\neq 1$.
\EndIf
\State \Return $l$, $f_{k}$ and the updated clusters.
\EndFunction
\end{algorithmic}
\end{table}

\begin{table}[!ht]
\begin{algorithmic}[1]
\Function{TAIL2HEAD}{$l,f_{K},( \hat{R}_{\textrm{s},K},\hat{R}_{\textrm{d},K}),( \hat{R}_{\textrm{s},1},\hat{R}_{\textrm{d},1})$}
\If {$( \hat{R}_{\textrm{s},K},\hat{R}_{\textrm{d},K} )=( \hat{R}_{\textrm{s},1},\hat{R}_{\textrm{d},1} )$}
    \State Combine the two clusters that contain $\textrm{T}_{K}$ and $\textrm{T}_{1}$ and set $l = l-1$ when $f_{K}=1$; Set $f_{K}=1$ and put $\textrm{T}_{K}$ into the cluster that contains $\textrm{T}_{1}$ when $f_{K}\neq 1$.
\Else \If {$(\hat{R}_{\textrm{s},K},\hat{R}_{\textrm{d},K}) = $REFLECTANCE$(1,\tau_{1},\vartheta_{K} + \Delta \theta - \pi/2)$}
        \State Set $f_{1}=1$; Put $\textrm{T}_{1}$ into the cluster that contains $\textrm{T}_{K}$ when $f_{K}=1$; Set $l=l+1$, $f_{K}=1$, create the $l$-th cluster as $\{\textrm{T}_{K},\textrm{T}_{1}\}$, and label it with reflectance $( \hat{R}_{\textrm{s},K},\hat{R}_{\textrm{d},K},\hat{\mathit{\Phi}}_{\textrm{re},K} )$ when $f_{K}\neq 1$.
    \EndIf
\EndIf
\State \Return $l$, $f_{1}$, $f_{K}$ and the updated clusters.
\EndFunction
\end{algorithmic}
\vspace{-1cm}
\end{table}
\linespread{2.0}

\subsection{Proof of Proposition 2}\label{Proof_P2}
Assume there is an $l$ satisfying the condition that $\Theta(\hat{x}_{\textrm{ue},m},\hat{y}_{\textrm{ue},m})$ falls within $\Omega_{l}$. In this sense, the LoS ray $y=\frac{\hat{y}_{\textrm{ue},m}}{\hat{x}_{\textrm{ue},m}}x$ intersects the $l$-th surface $y=a_{l}x + b_{l}$ at point
\begin{equation}
(x_{\textrm{ip}},y_{\textrm{ip}}) = \left(\frac{b_{l}\hat{x}_{\textrm{ue},m}}{\hat{y}_{\textrm{ue},m} - a_{l}\hat{x}_{\textrm{ue},m}},\frac{b_{l}\hat{y}_{\textrm{ue},m}}{\hat{y}_{\textrm{ue},m} - a_{l}\hat{x}_{\textrm{ue},m}}\right).
\label{intersection_point1}
\end{equation}
If the LoS path is blocked by surface $l$, then it can be inferred that $x_{\textrm{ip}}^{2}+y_{\textrm{ip}}^{2}<\hat{x}_{\textrm{ue},m}^{2}+\hat{y}_{\textrm{ue},m}^{2}$. That is to say,
\begin{equation}
\left|\frac{b_{l}}{y_{\textrm{ue},m} - a_{l}x_{\textrm{ue},m}}\right|<1.
\end{equation}
Since the LoS path exists if and only if it is not blocked by any reflective surfaces, Proposition 2 is naturally derived.

\vspace{-0.5cm}
\subsection{Proof of Proposition 3}\label{Proof_P3}
To ensure that the reflected ray arrives at the UE, two conditions must be satisfied. First, the mirror image $(x_{\textrm{ue},m}^{(l)},y_{\textrm{ue},m}^{(l)})$ should be no closer to the AP at the origin than the reflection point $(x_{\textrm{re},m}^{(l)},y_{\textrm{re},m}^{(l)})$, i.e.,
\begin{equation}
\left|\frac{b_{l}}{y_{\textrm{ue},m}^{(l)} -a_{l}x_{\textrm{ue},m}^{(l)}}\right|<1.
\end{equation}
Second, the reflected ray should not be blocked by other surfaces. As in (\ref{intersection_point1}), we calculate the intersection point of surface $y=a_{n}x+b_{n}$ and the reflected ray $y=\frac{y_{\textrm{re},m}^{(l)}-\hat{y}_{\textrm{ue},m}}{x_{\textrm{re},m}^{(l)}-\hat{x}_{\textrm{ue},m}}\left(x-\hat{x}_{\textrm{ue},m}\right)+\hat{y}_{\textrm{ue},m}$, which is given by
\begin{align}
&(x_{\textrm{ip}},y_{\textrm{ip}})= \notag \\
&\Big(\frac{\hat{x}_{\textrm{ue},m}y_{\textrm{re},m}^{(l)} - x_{\textrm{re},m}^{(l)}\hat{y}_{\textrm{ue},m} + b_{n}(x_{\textrm{re},m}^{(l)} - \hat{x}_{\textrm{ue},m})}{(y_{\textrm{re},m}^{(l)} - \hat{y}_{\textrm{ue},m})-a_{n}(x_{\textrm{re},m}^{(l)} - \hat{x}_{\textrm{ue},m})},\frac{a_{n}(\hat{x}_{\textrm{ue},m}y_{\textrm{re},m}^{(l)} - x_{\textrm{re},m}^{(l)}\hat{y}_{\textrm{ue},m}) + b_{n}(y_{\textrm{re},m}^{(l)} - \hat{y}_{\textrm{ue},m})}{(y_{\textrm{re},m}^{(l)} - \hat{y}_{\textrm{ue},m})-a_{n}(x_{\textrm{re},m}^{(l)} - \hat{x}_{\textrm{ue},m})} \Big).
\end{align}
If the reflected ray is blocked by surface $n$, then $\Theta\left(x_{\textrm{ip}},y_{\textrm{ip}}\right)\in \Omega_{n}$. Also, the intersection point $(x_{\textrm{ip}},y_{\textrm{ip}})$ must be located between $(\hat{x}_{\textrm{ue},m},\hat{y}_{\textrm{ue},m})$ and $(x_{\textrm{re},m}^{(l)},y_{\textrm{re},m}^{(l)})$, i.e., $(x_{\textrm{ip}}-\hat{x}_{\textrm{ue},m})^{2}+(y_{\textrm{ip}}-\hat{y}_{\textrm{ue},m})^{2}<(x_{\textrm{re},m}^{(l)}-\hat{x}_{\textrm{ue},m})^{2}+(y_{\textrm{re},m}^{(l)}-\hat{y}_{\textrm{ue},m})^{2}$ and $(x_{\textrm{ip}}-x_{\textrm{re},m}^{(l)})^{2}+(y_{\textrm{ip}}-y_{\textrm{re},m}^{(l)})^{2}<(x_{\textrm{re},m}^{(l)}-\hat{x}_{\textrm{ue},m})^{2}+(y_{\textrm{re},m}^{(l)}-\hat{y}_{\textrm{ue},m})^{2}$. These two inequalities can be rewritten as
\begin{align}
&\left|\frac{ a_{n}x_{\textrm{re},m}^{(l)} + b_{n} - y_{\textrm{re},m}^{(l)} }{ (y_{\textrm{re},m}^{(l)}-\hat{y}_{\textrm{ue},m}) - a_{n}(x_{\textrm{re},m}^{(l)}-\hat{x}_{\textrm{ue},m}) }\right|<1, \mbox{~and~} \notag \\
&\left|\frac{ a_{n}\hat{x}_{\textrm{ue},m} + b_{n} - \hat{y}_{\textrm{ue},m} }{ (y_{\textrm{re},m}^{(l)}-\hat{y}_{\textrm{ue},m}) - a_{n}(x_{\textrm{re},m}^{(l)}-\hat{x}_{\textrm{ue},m}) }\right|<1.
\end{align}
Combining these results, the proof is complete.

\bibliographystyle{IEEEtran}

\end{document}